\documentclass[authoryear,final,5p,twocolumn]{elsarticle}

\usepackage{amssymb}
\usepackage{amsthm}
\usepackage{amsmath}
\usepackage{lipsum}
\usepackage{array}
\usepackage{graphicx}
\usepackage{caption}
\usepackage{subcaption}
\usepackage{color}

\makeatletter
\def\ps@pprintTitle{%
   \let\@oddhead\@empty
   \let\@evenhead\@empty
   \let\@oddfoot\@empty
   \let\@evenfoot\@oddfoot
}
\makeatother
   
\begin{document}
   
\begin{frontmatter}

\title{Model-plant mismatch learning offset-free model predictive control\tnoteref{label1}}
\tnotetext[label1]{This paper was not presented at any IFAC 
meeting.}
\author[a]{Sang Hwan Son}\ead{zshson@gmail.com}
\author[b]{Jong Woo Kim}\ead{jong.w.kim@tu-berlin.de}
\author[c]{Tae Hoon Oh}\ead{rozenk@snu.ac.kr}
\author[c]{Jong Min Lee\corref{cor1}}\ead{jongmin@snu.ac.kr}
\address[a]{Texas A\&M Energy Institute, Texas A\&M University, College Station, TX 77845 USA}
\address[b]{Technische Universit{\"a}t Berlin, KIWI Biolab, Ackerstrasse 76, 13355 Berlin, Germany}
\address[c]{School of Chemical and Biological Engineering, Institute of Chemical Processes, Seoul National University, 1 Gwanak-ro, Gwanak-gu, Seoul 08826, Korea}
\cortext[cor1]{Corresponding author J.~M.~Lee. Tel. +82-02-880-1504. 
Fax +82-02-888-7295.}


\begin{abstract}
We propose model-plant mismatch learning offset-free model predictive control (MPC), which learns and applies the intrinsic model-plant mismatch, to effectively exploit the advantages of model-based and data-driven control strategies and overcome the limitations of each approach. In this study, the model-plant mismatch map on steady-state manifold in the controlled variable space is approximated via a general regression neural network from the steady-state data for each setpoint. Though the learned model-plant mismatch map can provide the information at the equilibrium point (i.e., setpoint), it cannot provide model-plant mismatch information during the transient state. Moreover, the intrinsic model-plant mismatch can vary due to system characteristics changes during operation. Therefore, we additionally apply a supplementary disturbance variable which is updated from the disturbance estimator based on the nominal offset-free MPC scheme. Then, the combined disturbance signal is applied to the target problem and finite-horizon optimal control problem of offset-free MPC to improve the prediction accuracy and closed-loop performance of the controller. By this, we can exploit both the learned model-plant mismatch information and the stabilizing property of the nominal disturbance estimator approach. The closed-loop simulation results demonstrate that the developed scheme can properly learn the intrinsic model-plant mismatch and efficiently improve the model-plant mismatch compensating performance in offset-free MPC. Moreover, we examine the robust asymptotic stability of the developed offset-free MPC scheme, which is known to be difficult to analyze in nominal offset-free MPC, by exploiting the learned model-plant mismatch information.
\end{abstract}

\end{frontmatter}

\section{Introduction}\label{sec1}

Model-based control strategies manipulate a system based on the prediction of the future state of a process using a known model. Model predictive control (MPC) which derives a finite-horizon optimal solution in a receding horizon manner is one of the most representative model-based approaches \citep*{38,34,32,31}. MPC can effectively derive a reliable solution based on the model, but the performance of MPC is directly related to the prediction accuracy of the model. Since model-plant mismatch and unmeasured disturbance always exist in real processes, MPC usually cannot achieve the optimal closed-loop performance \citep*{39,30}. Data-driven control strategies, which are mainly based on machine learning (ML) techniques, learn and utilize the system behavior such as dynamics or reward from real process data (e.g., reinforcement learning (RL) directly derives optimal control policy from the process data) \citep*{40,41,42,43}. Therefore, data-driven strategies do not indispensably require given models or prior assumptions about processes, and can implicitly manage uncertainties in processes. However, pure data-driven approaches without any prior knowledge of the process are often restricted because they require a large quantity of data, and exploratory policies during learning can cause system safety problems \citep*{33,8}.

Since model-based and data-driven control strategies are complementary, combining both strategies is an emerging area of research \citep*{35}. \cite*{1,2} derived reliable policies by approximating a nominal MPC policy through supervised learning methods such as guided policy search. \cite*{3} approximated the policy of a robust MPC by learning a neural network from the sampled data of robust MPC. \cite*{4,5} improved MPC performance by continuously updating a dynamic model, which was approximated with multilayer neural network from sampled data, and \cite*{6} improved MPC performance with a dynamic model, which was updated by sparse identification of nonlinear dynamics. \cite*{7} proposed learning both approximated MPC policy and system dynamics with a recurrent neural network and multilayer perceptron, respectively. \cite{8} proposed learning a direct compensatory control action for MPC that improves the closed-loop performance of the system by utilizing RL with the same performance measure of MPC. \cite*{44,45} proposed learning-based robust predictive control schemes that identifies prediction model and bounded uncertainty from process data and design robust MPC system.

These existing methods can be classified into two categories according to the object of learning: prediction model for MPC or policy of MPC. Each existing method presents a valuable direction for combining model-based and data-driven control strategies. However, it has not been studied yet that focuses on the complementarity between model-based and data-driven control strategies (i.e., the availability of previous knowledge of the process and the availability of process characteristics contained in data that cannot be reached with previous knowledge). Therefore, in this study, we propose a scheme that performs a predictive control based on a model built on previous knowledge of the process and simultaneously learns the model-plant mismatch from the process data and utilizes it for offset-free tracking. The proposed scheme is developed based on the offset-free MPC framework, which is one of the most representative model-plant mismatch compensating schemes in MPC field \citep*{17,13,15}.

Specifically, in the proposed scheme, the general regression neural network (GRNN) proposed in \cite*{19} is constructed to approximate the model-plant mismatch map on steady-state manifold in controlled variable space from the data. Then, the disturbance value derived from the approximated model-plant mismatch map is applied to the target problem and optimal control problem of offset-free MPC. However, since the learned model-plant mismatch map can only provide the information at the equilibrium point, the supplementary disturbance signal, which is updated from the disturbance estimator, is also applied to handle the transient state of the process and the change in the intrinsic model-plant mismatch due to system characteristics change during operation. In this way, the proposed scheme effectively improves the closed-loop tracking performance of offset-free MPC by exploiting both the learned model-plant mismatch information and the stabilizing property of the disturbance estimator. Moreover, we examine the robust asymptotic stability of the proposed offset-free MPC scheme by exploiting the learned model-plant mismatch information, which has not been done rigorously in the existing offset-free MPC studies due to the difficulty in handling the combined system consisting of a disturbance estimator, a target problem, and an optimal control problem \citep*{20}.

The rest of this paper is organized as follows. We introduce the standard offset-free MPC formulation and offset-free tracking condition in Section~\ref{sec2}. In Section~\ref{sec3}, we develop the model-plant mismatch learning offset-free MPC and examine the robust asymptotic stability of the proposed scheme. In Section~\ref{sec4}, we present the numerical examples to demonstrate the efficacy of the proposed model-plant mismatch learning offset-free MPC compared to the nominal offset-free MPC under various operating conditions. In Section~\ref{sec5}, we conclude with a few important remarks.
 
\section{Offset-free MPC: Disturbance estimator approach}\label{sec2}
We present the standard linear offset-free MPC framework in \cite*{18,14}.

\subsection{Preliminaries}
Consider the discrete time-invariant plant in the form:
\begin{align}\label{eq1}
\begin{cases} x_p(k+1)=f_p( x_p(k), u(k)) \\
y_p(k)=g_p(x_p(k)) \\ z_p(k)=Hy_p(k) \end{cases} 
\end{align}
with constraints
\begin{align}\label{eq2}
u\in{\mathcal{U}},\; x_p\in{\mathcal{X}}
\end{align}
where $x_p \in\mathbb{R}^{n_{x_p}}$, $u\in\mathbb{R}^{n_u}$, $y_p\in\mathbb{R}^{n_{y}}$, and $z_p\in\mathbb{R}^{n_{z}}$ are the state, input, output, and controlled variables, respectively. Without loss of generality, the matrix $H$ is assumed to have full row rank. $\mathcal{U}$ and $\mathcal{X}$ are constraint sets presented as compact polyhedral regions.

The objective of offset-free linear MPC is to make the plant controlled variables $z_p$ track the reference with the linear time-invariant system model of (1).
\begin{align}\label{eq3}
\begin{cases} x(k+1)=A x(k)+B u(k) \\ y(k)=C x (k) \end{cases} 
\end{align}
where $x\in\mathbb{R}^{n_x}$ and $y\in\mathbb{R}^{n_y}$ are the model state and output, respectively. The reference $r(k)$ is assumed to converge to a constant value $r_\infty$ as $k \rightarrow \infty$. The pair (A, B)  and (C, A) are assumed to be controllable and observable, respectively.

\subsection{Disturbance estimator and optimal control problem}
The most standard method to compensate for the mismatch between the plant in (1) and the model in (3), and achieve offset-free reference tracking at steady state is to augment the plant model with additional integrating state so-called disturbance as in (4).
\begin{align}\label{eq4}
\begin{cases} x(k+1)=A x(k)+B u(k)+B_d d(k) \\ d(k+1)=d(k) \\ y(k)=C x (k)+C_d d(k) \end{cases} 
\end{align}
where $d\in\mathbb{R}^{n_d}$ is the disturbance, and $B_d\in\mathbb{R}^{n_x \times n_d}$ and $C_d\in\mathbb{R}^{n_y \times n_d}$ are disturbance model matrices.\\
\quad\\
\textbf{Proposition 1}. The augmented system is observable if and only if the pair (C, A) is observable and (5) holds \citep{18}.
\begin{align}\label{eq5}
{\mathrm{rank}} \begin{bmatrix} A-I & B_d \\ C & C_d \end{bmatrix} =n_x+n_d.
\end{align}

The appropriate matrices $B_d$ and $C_d$ that satisfy the condition in (5) can exist if and only if the number of disturbances $n_d$ is equal or smaller than that of the measured outputs $n_y$, $n_d \leq n_y$ \citep*{16}.\\

In the assumption that $B_d$ and $C_d$ are chosen to satisfy the condition in \textbf{Proposition 1}, the state and disturbance estimator is designed as in (6).
\begin{align}\label{eq6}
\begin{bmatrix} \hat{x}(k+1) \\ \hat{d}(k+1) \end{bmatrix}&=
\begin{bmatrix} A & B_d \\ 0 & I \end{bmatrix} \begin{bmatrix} \hat{x}(k) \\ \hat{d}(k) \end{bmatrix}+ \begin{bmatrix} B \\ 0 \end{bmatrix} u(k) \nonumber\\
& + \begin{bmatrix} L_x \\ L_d \end{bmatrix} (-y_p(k)+\begin{bmatrix} C & C_d \end{bmatrix} \begin{bmatrix} \hat{x}(k) \\ \hat{d}(k) \end{bmatrix})
\end{align}
where $L_x\in\mathbb{R}^{n_x \times n_y}$ and $L_d\in\mathbb{R}^{n_d \times n_y}$ are the estimator gains for state and disturbance, respectively, which are chosen to make the estimator stable.

The following finite-horizon optimal control problem P is solved in receding horizon manner. 
\begin{align}
\mathrm{P}:\;J^{*}=\underset{\small{u_0,\cdots,u_{N-1}}}{\mathrm{min}}&\; \phi_t(x_N)+\sum _{ i=0 }^{ N-1 }{\phi(x_i,u_i)} \nonumber\\
\mathrm{s.t.}\quad &x_{0}=\hat{x},\; d=\hat{d}\nonumber\\
&x_{i+1}=Ax_{i}+Bu_{i}+B_d d\nonumber\\
&u_{i}\in{\mathcal{U}},\; x_{i+1}\in{\mathcal{X}},\; x_{N}\in {\mathcal{X}}_t \nonumber\\
&i=0,\dots ,N-1\nonumber
\end{align}
with target state $\bar{x}$ and input $\bar{u}$ are given by
\begin{align}\label{eq7}
\begin{bmatrix} A-I & B \\ HC & 0 \end{bmatrix}
\begin{bmatrix} \bar{x} \\ \bar{u} \end{bmatrix}
= \begin{bmatrix} -B_d \hat{d}(k) \\ r-HC_d \hat{d}(k) \end{bmatrix}
\end{align}
where $\phi(x_i,u_i):=|x_i-\bar{x}|^2_{Q_x}+|u_i-\bar{u}|^2_{Q_u}$ and $\phi_t(x_N):=|x_N-\bar{x}|^2_{Q_x^N}$ denote the single stage cost and terminal cost with $|v|^2_Q:=v^\top Qv$, respectively, ${\mathcal{X}}_t$ is the terminal constraint set, $Q_x\in{\mathbb{R}}^{n_x}$ and $Q_u\in{\mathbb{R}}^{n_u}$ are weighting matrices with diagonal form, and $r\in\mathbb{R}^{n_z}$ denotes the reference.

\subsection{Offset-free tracking condition}

By rearranging (6), we can see that the disturbance estimator satisfies (8) at steady state.
\begin{align}\label{eq8}
\begin{bmatrix} A-I+L_x C & B \\ L_d C & 0 \end{bmatrix}
\begin{bmatrix} \hat{x}_\infty \\ u_\infty \end{bmatrix}
= \begin{bmatrix} L_x y_{p,\infty} -(B_d+L_x C_d) \hat{d}_\infty \\ L_d y_{p,\infty}-L_d C_d \hat{d}_\infty \end{bmatrix}
\end{align}
where $\infty$ denotes steady-state values. 

Let $\kappa_{un}$ denote the unconstrained optimal gain for the state of P. Assume that the MPC problem in P is feasible for all $k\in\mathbb{N}^+$ and unconstrained for $k\geq j$ with $j\in\mathbb{N}^+$. Then, $\kappa_{un}$ is optimal and feasible at steady state, and thus, the steady-state input $u_\infty$ and target input $\bar{u}_\infty$ satisfy (9).
\begin{align}\label{eq9}
u_\infty -\bar{u}_\infty=\kappa_{un}(\hat{x}_\infty-\bar{x}_\infty ).
\end{align}

Let $e_{y,\infty}$ and $e_{z,\infty}$ denote the prediction error of the steady-state output and the offset of the controlled variable as in (10) and (11).
\begin{align}
&e_{y,\infty} :=y_{p,\infty}-C\hat{x}_\infty-C_d \hat{d}_\infty.\label{eq10} \\
&e_{z,\infty} := Hy_{p,\infty}-r_\infty.\label{eq11}
\end{align}
Then, the offset-free tracking condition can be represented as in \textbf{Proposition 2}.

\quad\\
\textbf{Proposition 2}. If the following condition in (12) is satisfied, then offset-free tracking is achieved at steady state \citep{18}.
\begin{align}\label{eq12}
\mathcal{N}(L_d) \subseteq \mathcal{N}(H(I-C(I-A-B\kappa_{un})^{-1}L_x)
\end{align}
where $\mathcal{N}$ represents the null space.\\
\quad\\
\textbf{Proof}. By combining and rearranging (7)--(11), we can derive (13).
\begin{align}\label{eq13}
\begin{bmatrix} L_d \\ H(I-C(I-A-B\kappa_{un})^{-1}L_x) \end{bmatrix} e_{y,\infty} = \begin{bmatrix} 0 \\ I \end{bmatrix}e_{z,\infty}.
\end{align}
We can see that (14) should be satisfied to achieve offset-free tracking at steady state, i.e., $e_{z,\infty}=0$, for all $e_{y,\infty}$ satisfying $L_de_{y,\infty}=0$.
\begin{align}\label{eq14}
H(I-C(I-A-B\kappa_{un})^{-1}L_x) e_{y,\infty}=0.
\end{align}
Then, this condition can be reformulated into the null space condition in (12).\qed\\
\quad\\
\textbf{Remark 1}. 
To construct $L_x$ and $L_d$ satisfying the condition in \textbf{Proposition 2}, \cite{14} suggested the estimator gain structure in (15).
\begin{align}
& \begin{bmatrix} L_x \\ L_d \end{bmatrix}=\begin{bmatrix} L_x^0 \\ 0 \end{bmatrix} +\begin{bmatrix} \bar{L}_x \\ \bar{L}_d \end{bmatrix} H(I-C{e_{\kappa_x,\infty}}^{-1}L_x^0) \label{eq15}\\
& e_{\kappa_x,\infty}:=I-A-B\kappa_{un} \nonumber.
\end{align}
When the number of disturbance variables $n_d$ and the number of measured outputs $n_y$ are identical, $n_d = n_y$, by \textit{Proposition 2} in \cite{14}, $L_d$ is nonsingular. In this case, $e_{y,\infty}$ naturally becomes 0 at steady state, therefore, a simple gain structure, where $\bar{L}_x$ and $\bar{L}_d$ are chosen to stabilize the estimator in (6) with $L_x^0=0$, can satisfy the offset-free tracking condition in \textbf{Proposition 2}. In the case of $n_d < n_y$, $L_x^0$, $\bar{L}_x$, and $\bar{L}_d$ can be constructed according to the procedures suggested in \textit{Algorithms 4.2} and \textit{4.3} in \cite{14}.

\section{Model-plant mismatch learning offset-free MPC}\label{sec3}

Though the nominal offset-free MPC in Section~\ref{sec2} can compensate for the model-plant mismatch and have been implemented various processes \cite{46,47,48}, it basically estimates the proper disturbance from the occurred measurement error. Therefore, there exists some delay in estimating the proper disturbance value and compensating for the model-plant mismatch. Moreover, since the model-plant mismatch varies for each system state, the estimated disturbance at one state point is not available for another state point. Due to these limitations, the nominal offset-free MPC cannot properly handle the model-plant mismatch in the transition state, such as setpoint changes, and considerable closed-loop performance degradation can occur.

To overcome this limitation, we propose learning the intrinsic model-plant mismatch from the past estimated steady-state disturbance data and applying the learned model-plant mismatch to the disturbance estimator, target problem, and model-based finite-horizon optimal control problem to improve the closed-loop performance of the control system.

\subsection{Model-plant mismatch learning} 

By applying the estimated disturbance $\hat{d}$ from (6) to P, we can compensate the effect of the model-plant mismatch on future states and improve the prediction accuracy. This compensating disturbance signal value is intrinsically given for each plant state, model state, and input triple $(x_p,x,u)$ with proper disturbance model matrices, $B_d$ and $C_d$. Therefore, we can consider this disturbance value for each $(x_p,x,u)$ as the intrinsic model-plant mismatch.

If we can obtain the entire model-plant mismatch map for every $(x_p,x,u)$, we can accurately predict the real plant output by compensating for all the model-plant mismatch for future state. However, since obtaining the entire model-plant mismatch map is consequently identical to obtaining the exact plant dynamics, it requires massive data and computation.

Therefore, we propose to learn and utilize a reduced model-plant mismatch map only on steady-state manifold in the controlled variable space. This reduced model-plant mismatch map is a tiny part of the entire model-plant mismatch map. Therefore, we can derive this reduced model-plant mismatch map with considerably smaller quantities of data and computation than those required for the entire map.
 
\quad\\
\textbf{Lemma 1}. Assume that the estimator in (6) is stable. Then, the matrix $\begin{bmatrix} A-I+L_x C & B_d+L_x C_d \\ L_d C & L_d C_d \end{bmatrix}$ is nonsingular.\\
\quad\\
\textbf{Proof}. Rearranging (6) follows (16).
\begin{align}\label{eq16}
\begin{bmatrix} \hat{x}(k+1) \\ \hat{d}(k+1) \end{bmatrix}= &
\begin{bmatrix} A+L_x C & B_d+L_x C_d \\ L_d C & I+L_d C_d \end{bmatrix} \begin{bmatrix} \hat{x}(k) \\ \hat{d}(k) \end{bmatrix} \nonumber\\
&+ \begin{bmatrix} B \\ 0 \end{bmatrix} u(k) - \begin{bmatrix} L_x \\ L_d \end{bmatrix} y_p(k).
\end{align}
Since we assumed that the estimator is stable, it has no eigenvalue with 1. Therefore, the following (17) holds.
\begin{align}\label{eq17}
{\mathrm{det}} \left( \begin{bmatrix} A+L_x C & B_d+L_x C_d \\ L_d C & I+L_d C_d \end{bmatrix} -I \right) \neq 0.
\end{align}
Thus, the matrix $\begin{bmatrix} A-I+L_x C & B_d+L_x C_d \\ L_d C & L_d C_d \end{bmatrix}$ is nonsingular.\qed\\

With \textbf{Lemma 1}, we can show the existence of a state and disturbance estimates pair at steady state for each setpoint.

\quad\\
\textbf{Theorem 1}. If a setpoint $\bar{r}$ is physically reachable, i.e., there exists a steady-state plant output and input pair ($y_{p,\infty}$, $u_\infty$) that achieves $z_{p,\infty}=\bar{r}$, then, an estimated model state and disturbance pair ($\hat{x}_\infty$, $\hat{d}_\infty$) at that steady state always exists.\\
\quad\\
\textbf{Proof}. If a setpoint $\bar{r}$ is achievable, we can see (18) and (19) are satisfied at that steady state from the estimator in (6).
\begin{align}
&\hat{x}_\infty=(A+L_x C) \hat{x}_\infty +(B_d + L_x C_d) \hat{d}_\infty \nonumber\\
& \qquad+ B u_\infty - L_x y_{p,\infty} \label{eq18}\\
&\hat{d}_\infty=L_d C \hat{x}_\infty +(I+ L_d C_d) \hat{d}_\infty - L_d y_{p,\infty}\label{eq19}
\end{align}
where $y_{p,\infty}$ denotes the steady-state plant output that satisfies $\bar{r} =Hy_{p,\infty}$. $u_\infty$ denotes the steady-state input, and $\hat{x}_\infty$ and $\hat{d}_\infty$ denote the state and disturbance estimates at steady state, respectively. By rearranging (18) and (19), we can derive (20).
\begin{align}
\begin{bmatrix} A-I+L_x C & B_d+L_x C_d \\ L_d C & L_d C_d \end{bmatrix}
\begin{bmatrix} \hat{x}_\infty \\ \hat{d}_\infty \end{bmatrix}
= \begin{bmatrix} L_x y_{p,\infty} -B u_\infty \\ L_d y_{p,\infty} \end{bmatrix}.\label{eq20}
\end{align}
Since the matrix $\begin{bmatrix} A-I+L_x C & B_d+L_x C_d \\ L_d C & L_d C_d \end{bmatrix}$ is nonsingular from \textbf{Lemma 1}, we can see the pair ($\hat{x}_\infty$, $\hat{d}_\infty$) always exists whenever a steady-state plant output and input pair ($y_{p,\infty}$, $u_\infty$) achieving $\bar{r}$ exists. \qed\\

By rearranging (20), the ($\hat{x}_\infty$, $\hat{d}_\infty$) pair can be derived directly from the ($y_{p,\infty}$, $u_\infty$) pair.
\begin{align}\label{eq21}
\begin{bmatrix} \hat{x}_\infty \\ \hat{d}_\infty \end{bmatrix}
= \begin{bmatrix} A-I+L_x C & B_d+L_x C_d \\ L_d C & L_d C_d \end{bmatrix}^{-1} \begin{bmatrix} L_x & -B \\ L_d & 0 \end{bmatrix} \begin{bmatrix} y_{p,\infty} \\ u_\infty \end{bmatrix}.
\end{align}
\textbf{Remark 2}. We can see that the ($\hat{x}_\infty$, $\hat{d}_\infty$) pair is derived from the ($y_{p,\infty}$, $u_\infty$) pair as in (21). Therefore, the uniqueness of the ($\hat{x}_\infty$, $\hat{d}_\infty$) pair for each setpoint is not confirmed when more than one ($y_{p,\infty}$, $u_\infty$) pair can achieve the setpoint $\bar{r}$. In this study, we focus on the case where only one ($y_{p,\infty}$, $u_\infty$) pair achieves each $\bar{r}$, and thus, the uniqueness of ($\hat{x}_\infty$, $\hat{d}_\infty$) for each $\bar{r}$ is ensured.\\

Since the existence and uniqueness of ($\hat{x}_\infty$, $\hat{d}_\infty$) for each $\bar{r}$ are ensured by \textbf{Theorem 1} and \textbf{Remark 2}, we can define an intrinsic relation between $\bar{r}$ and $\hat{d}_\infty$ as (22).
\begin{align}\label{eq22}
\hat{d}_\infty =f_d (\bar{r} ).
\end{align}
The function $f_d\: :\:\mathbb{R}^{n_z}\rightarrow\mathbb{R}^{n_d}$ implies the intrinsic model-plant mismatch of the system.

We approximate the function $f_d(\cdot)$ in (22) with GRNN from the estimated disturbance data. GRNN is a variation of radial basis function (RBF) based neural networks for non-parametric regression proposed by \cite{19}. GRNN can be interpreted as a normalized RBF network with hidden units centered at every training sample. The predicted output $o(i)$ from the input $i$ by GRNN is a weighted average of outputs for the training set:
\begin{align}\label{eq23}
o(i)=\frac{\sum_{s=1}^{N_s}o_s \omega(i,i_s)}{\sum_{s=1}^{N_s}\omega(i,i_s)}
\end{align}
where $N_s$ is the number of training samples, and $\omega(i,i_s)$ denotes the weight. Each weight is an RBF output that is the exponential of the negatively scaled distance between the new pattern and each given training pattern:
\begin{align}\label{eq24}
\omega(i,i_s)=e^{-(i-i_s)^\top (i-i_s)/2\sigma^2}
\end{align}
where $\sigma$ is the smoothing factor which represents the width of the RBF.

\quad\\
\textbf{Remark 3}. The GRNN is a single-pass learning network with no training parameters while the backpropagation neural network (BPNN) needs forward and backward pass training. The only adjustable parameter in GRNN is the smoothing factor $\rho$ \citep*{21}.\\

By \textbf{Remark 3}, GRNN needs significantly less time for training than BPNN. By this notable advantage of rapid training, GRNN is suitable for on-line systems or systems that require minimal computation \citep*{22}.

However, since the number of neurons in the hidden layer is equal to the number of training samples,  the size of the GRNN can be large. We overcome this limitation by using only a few recent data in the past time window for learning. This strategy is reasonable, since the characteristics of the system can change during operation. 

\subsection{Application of learned model-plant mismatch}

In this section, we introduce how to utilize the learned model-plant mismatch value $\hat{d}^{\ell}$ in (25).
\begin{align}\label{eq25}
\hat{d}^{\ell} =\hat{f}_d (\bar{r})
\end{align}
where $\hat{f}_d(\cdot)$ is the approximated function of $f_d(\cdot)$ in (22) with GRNN. 

Since, this learned model-plant mismatch map can only provide the information at the equilibrium point, it cannot provide model-plant mismatch information during the transient state. Therefore, we propose incorporating the learned model-plant mismatch into the nominal disturbance estimator to exploit the stabilizing property. For this, we introduce an additional supplementary disturbance variable $\hat{d}^{s}$ that is continually updated by a revised disturbance estimator: 
\begin{align}\label{eq26}
&\!\!\!\begin{bmatrix} \hat{x}^{\ell,s}(k+1) \\ \hat{d}^{s}(k+1) \end{bmatrix}=
\begin{bmatrix} A & B_d \\ 0 & I \end{bmatrix} \begin{bmatrix} \hat{x}^{\ell,s}(k) \\ \hat{d}^{s}(k) \end{bmatrix} + \begin{bmatrix} B \\ 0 \end{bmatrix} u(k) + \begin{bmatrix} B_d \\ 0 \end{bmatrix} \hat{d}^{\ell}(k) \nonumber\\
&+ \begin{bmatrix} L_x \\ L_d \end{bmatrix}(-y_p(k)+C\hat{x}^{\ell,s}(k)+C_d(\hat{d}^{\ell}(k)+ \hat{d}^{s}(k))).
\end{align}
where $\hat{x}^{\ell,s}$ denotes the state estimate considering the learned model-plant mismatch $\hat{d}^{\ell}$ and the supplementary disturbance $\hat{d}^{s}$. 

\quad\\
\textbf{Theorem 2}. If $L_x$ and $L_d$ are chosen to make the nominal disturbance estimator in (6) stable, then the proposed disturbance estimator in (26) is also stable with the same $L_x$ and $L_d$.\\
\quad\\
\textbf{Proof}. Rearranging (26) follows (27).
\begin{align}\label{eq27}
&\begin{bmatrix} \hat{x}^{\ell,s}(k+1) \\ \hat{d}^{s}(k+1) \end{bmatrix}=
\begin{bmatrix} A+L_x C & B_d+L_x C_d \\ L_d C & I+L_d C_d \end{bmatrix} \begin{bmatrix} \hat{x}^{\ell,s}(k) \\ \hat{d}^{s}(k) \end{bmatrix} \nonumber\\
&\quad+\begin{bmatrix} B \\ 0 \end{bmatrix} u(k) -\begin{bmatrix} L_x \\ L_d \end{bmatrix} y_p(k)+\begin{bmatrix} B_d+L_x C_d \\ L_d C_d \end{bmatrix} \hat{d}^{\ell}(k).
\end{align}

Let $x$ and $d^s$ denote the exact model state and supplementary disturbance:
\begin{align}\label{eq28}
\begin{bmatrix} x(k+1) \\ d^s(k+1) \end{bmatrix}=\begin{bmatrix} A & B_d \\ 0 & I \end{bmatrix} \begin{bmatrix} x(k) \\ d^s(k) \end{bmatrix}+\begin{bmatrix} B \\ 0 \end{bmatrix} u(k).
\end{align}
These exact values satisfy (29). 
\begin{align}\label{eq29}
y_p(k)=Cx(k)+C_d(d^s(k)+\hat{d}^\ell(k)).
\end{align}
By substituting (29) into (27) and subtracting (27) from (28), we obtain the error dynamics in Eq.(30).
\begin{align}\label{eq30}
\begin{bmatrix} e_{\hat{x}}(k+1) \\ e_{\hat{d}^s}(k+1) \end{bmatrix}&=
\begin{bmatrix} A+L_x C & B_d+L_x C_d \\ L_d C & I+L_d C_d \end{bmatrix} \begin{bmatrix} e_{\hat{x}}(k) \\ e_{\hat{d}^s}(k) \end{bmatrix}
\end{align}
where $e_{\hat{x}}:=x-\hat{x}^{\ell,s}$ and $e_{\hat{d}^s}:=d^s-\hat{d}^s$.

Since the matrix $\begin{bmatrix} A+L_x C & B_d+L_x C_d \\ L_d C & I+L_d C_d \end{bmatrix}$ is the same as that of the nominal disturbance estimator in (6), we can see that the proposed estimator is stable with the same $L_x$ and $L_d$ of the nominal estimator.\qed
\quad\\

Then, $\hat{x}^{\ell,s}$ and the combined disturbance $\hat{d}^{\ell,s}$ in (31) are applied to the finite-horizon optimal control problem P and the target calculation in (7).
\begin{align}\label{eq31}
\hat{d}^{\ell,s}=\hat{d}^{\ell}+\hat{d}^{s}.
\end{align}

\quad\\
\textbf{Remark 4}. Even when $\hat{d}^{\ell}$ is not properly learned by GRNN or the intrinsic model-plant relation alters due to system characteristic changes, the supplementary signal $\hat{d}^{s}$ from the disturbance estimator in (27) compensates for the model-plant mismatch and makes the controller achieve the offset-free tracking property.
\quad\\

The zero steady-state offset property of the proposed model-plant mismatch learning offset-free MPC framework is proven in \textbf{Theorem 3}.

\quad\\
\textbf{Theorem 3}. Assume that the closed-loop system of the proposed framework with the estimator gains $L_x$ and $L_d$ constructed as in \textbf{Remark 1} converges to the steady state: $\hat{x}^{\ell,s}(k)\rightarrow\hat{x}_\infty^{\ell,s}$, $\hat{d}^{\ell,s}(k)\rightarrow\hat{d}_\infty^{\ell,s}$, and $y_{p}(k)\rightarrow y_{p,\infty}$ as $k\rightarrow\infty$. Then, $Hy_{p}(k)\rightarrow \bar{r}$ as $k\rightarrow\infty$.\\
\quad\\
\textbf{Proof}. By rearranging (27), we obtain (32) and (33) at steady state.
\begin{align}
&\hat{x}_\infty^{\ell,s}=A\hat{x}_\infty^{\ell,s}+Bu_\infty^{\ell,s}+B_d\hat{d}_\infty^{\ell,s}-L_xe_{y,\infty}^{\ell,s}\label{eq32}\\
&0=L_de_{y,\infty}^{\ell,s}\label{eq33}
\end{align}
where $e_{y,\infty}^{\ell,s}$ denotes the output reconstruction error of P$^{\ell,s}$ at steady state:
\begin{align}\label{eq34}
e_{y,\infty}^{\ell,s}:=y_{p,\infty}-C\hat{x}^{\ell,s}_{\infty}-C_d\hat{d}^{\ell,s}_{\infty}.
\end{align}
From the target calculation in (7), we can derive (35) and (36).
\begin{align}
&\bar{x}^{\ell,s}_\infty=A\bar{x}^{\ell,s}_\infty+B\bar{u}^{\ell,s}_\infty+B_d\hat{d}^{\ell,s}_\infty\label{eq35}\\
&\bar{r}=HC\bar{x}^{\ell,s}_\infty+HC_d\hat{d}^{\ell,s}_\infty.\label{eq36}
\end{align}
The steady-state input can be derived from the unconstrained optimal gain $\kappa_{un}$ of P:
\begin{align}\label{eq37}
u^{\ell,s}_\infty-\bar{u}^{\ell,s}_\infty=\kappa_{un}(\hat{x}^{\ell,s}_\infty-\bar{x}^{\ell,s}_\infty).
\end{align}
Subtracting (35) from (32) and substituting into (37) yields
\begin{align}\label{eq38}
\hat{x}^{\ell,s}_\infty-\bar{x}^{\ell,s}_\infty=-(I-A-B\kappa_{un})^{-1}L_xe_{y,\infty}^{\ell,s}.
\end{align}

Now, let $e_{z,\infty}^{\ell,s}$ denote the offset vector of the controlled variables at steady state:
\begin{align}\label{eq39}
e_{z,\infty}^{\ell,s}:=Hy_{p,\infty}-\bar{r}.
\end{align}
Substituting (36) into (39) and rearranging yields
\begin{align}\label{eq40}
e_{z,\infty}^{\ell,s}=H(y_{p,\infty}-C\hat{x}^{\ell,s}_\infty-C\hat{d}^{\ell,s}_\infty+C(\hat{x}^{\ell,s}_\infty-\bar{x}^{\ell,s}_\infty)).
\end{align}
Then, substituting (34) and (38) into (40) yields
\begin{align}\label{eq41}
e_{z,\infty}^{\ell,s}=H[I-C(I-A-B\kappa_{un})^{-1}L_x]e_{y,\infty}^{\ell,s}.
\end{align}
Finally, combining and rearranging (33) and (41), we obtain
\begin{align}\label{eq42}
\begin{bmatrix} L_d \\ H(I-C(I-A-B\kappa_{un})^{-1}L_x) \end{bmatrix} e_{y,\infty}^{\ell,s} = \begin{bmatrix} 0 \\ I \end{bmatrix}e_{z,\infty}^{\ell,s}.
\end{align}

Since the coefficient matrices in (42) are identical to those of (13), the same null space condition for nominal offset-free MPC in (12) is required for $e_{z,\infty}^{\ell,s}\rightarrow 0$, which can be satisfied by $L_x$ and $L_d$ constructed as in \textbf{Remark 1}. \qed\\

\subsection{Robust asymptotic stability of model-plant mismatch learning offset-free MPC}

To show the closed-loop asymptotic stability of offset-free MPC, we have to examine the closed-loop behavior of the combined system consisting of a disturbance estimator, a target problem, and an optimal control problem. Since it is known to be difficult to examine this closed-loop behavior of the combined system, almost all offset-free MPC schemes only show the offset-free tracking property when the stable equilibrium has been reached \citep{20,49}.

The main difficulty associated with this problem is that the nominal offset-free MPC cannot specify an equilibrium point and Lyapunov function candidate until the system reaches a steady state, since the target state and the target input in P are continually updated through the target problem with the estimated $\hat{d}$. However, in the proposed model-plant mismatch learning offset-free MPC framework, we can specify the equilibrium point and the Lyapunov function candidate using the disturbance value derived from the learned model-plant mismatch.

With the specified equilibrium point and Lyapunov function candidate, we examine the closed-loop robust asymptotic stability of model-plant mismatch learning offset-free MPC based on the framework in \cite*{23,24} which shows the closed-loop robust asymptotic stability of the combined system consisting of a estimator and an optimal control problem.

\cite{23} defined robust asymptotic stability as input-to-state stability (ISS) \citep*{25} on a robust positive invariant set ($\mathbf{v}$ represents the sequence of $v(k)$ for $k\in \mathbb{I}_{\geq 0}$ where $\mathbb{I}_{\geq 0}$ represents the set of nonnegative integers, $||\mathbf{v}||$ represents the supremum norm $||\mathbf{v}||=\mathrm{sup}_{k\geq0}|v(k)|$, and $|v|$ represents the vector 2-norm $|v|=(v^\top v)^{1/2}$).\\
\quad\\
\textbf{Definition 1}. (Robust positive invariance) Let $\Xi$ compact metric space. A set $\mathcal{O}\subseteq\Xi$ is said to be robust positive invariant for a perturbed system $\xi^+\in f_m(\xi,e)$ with perturbation $e$ if there exists some $\delta_e>0$ such that $f_m(\xi,e)\subseteq\mathcal{O}$ for all $\xi\in\mathcal{O}$ and $\mathbf{e}$ satisfying $||\mathbf{e}||\leq\delta_e$ \citep{23}.\\
\quad\\
\textbf{Definition 2}. (Robust asymptotic stability) The equilibrium point $\bar{\xi}$ of a perturbed system $\xi^+\in f_m(\xi,e)$ is robustly asymptotically stable in $\mathcal{O}$ if there exists some $\delta_e>0$ such that for all perturbation sequences $\mathbf{e}$, $||\mathbf{e}||<\delta_e$, $\mathcal{O}$ is robust invariant and there exists a class $\mathcal{KL}$ function $\beta(\cdot)$ and a class $\mathcal{K}$ function $\sigma(\cdot)$ satisfying
\begin{align}\label{eq43}
|\xi^+_k(\xi,\mathbf{e})-\bar{\xi}|\leq \beta(|\xi-\bar{\xi}|,k)+\sigma(||\mathbf{e}||)
\end{align}
for each $\xi\in\mathcal{O}$ and for all $k\in\mathbb{I}_{\geq 0}$ where $\xi^+_k(\xi,\mathbf{e})$ is the open-loop solution of the perturbed system for given time step $k$ \citep{23}.
\quad\\

Then, \cite{24} showed the robust asymptotic stability of the equilibrium point of the combined system by establishing that the value function of the finite-horizon optimal control problem is an ISS Lyapunov function for the combined system and applying \textbf{Proposition 3}.\\
\quad\\
\textbf{Definition 3}. (ISS Lyapunov function) $V(\cdot)$ is an ISS Lyapunov function in a robust positive invariant set $\mathcal{O}$ for a perturbed system $\xi^{+}\in f_m(\xi,e)$ with the equilibrium point $\bar{\xi}$ if there exists $\delta_e>0$, class $\mathcal{K}_\infty$ functions $\alpha_1(\cdot),\alpha_2(\cdot),\alpha_3(\cdot)$, and class $\mathcal{K}$ function $\sigma_e(\cdot)$ such that for all $\xi\in\mathcal{O}$ and $||\mathbf{e}||\leq \delta_e$ which satisfies (44) and (45) \citep*{26}.
\begin{align}
&\alpha_1(|\xi-\bar{\xi}|)\leq V(\xi-\bar{\xi}) \leq\alpha_2(|\xi-\bar{\xi}|)\label{eq44}\\
\underset{\xi^{+}\in f_m(\xi,e)}{\mathrm{sup}} &V(\xi^+-\bar{\xi})\leq V(\xi-\bar{\xi})-\alpha_3(|\xi-\bar{\xi}|)+\sigma_e(||\mathbf{e}||).\label{eq45}
\end{align}
\quad\\
\textbf{Proposition 3}. If there exists an ISS Lyapunov function for the perturbed system $\xi^+=f_m(\xi,e)$ with the equilibrium point $\bar{\xi}$ for all $||\mathbf{e}||\leq\delta_e$ for some $\delta_e>0$ on a robust positive invariant set $\mathcal{O}$, then the equilibrium point $\bar{\xi}$ is robustly asymptotically stable in $\mathcal{O}$ for all $||\mathbf{e}||\leq\delta_e$ \citep{24}.\\
\quad\\
\textbf{Proposition 4}. Let $\Gamma$ be a compact metric space and $g:\Gamma\rightarrow\mathbb{R}^n$ be continuous. Then, there exists a class $\mathcal{K}$ function $\sigma(\cdot)$ such that $|g(p)-g(q)|\leq \sigma(|p-q|)$ for all $p,q\in\Gamma$. The proof is provided in \textit{Appendix A}.

\subsubsection{Equilibrium point and Lyapunov function candidate}

We examine the closed-loop behavior of the combined system under the assumption that the approximated model-plant mismatch function $\hat{f}_d(\cdot)$ in (25) is completely learned.

Let $\hat{f}_d^*(\cdot)$ denote this completely learned model-plant mismatch map and $\hat{d}^{\ell^*}(\bar{r})$ denote the disturbance value corresponding to the set-point $\bar{r}$ derived from $\hat{f}_d^*(\cdot)$. Then, we can consider the target state $\bar{x}^{\ell^*}$ and target input $\bar{u}^{\ell^*}$ derived by substituting $\hat{d}^{\ell^*}$ into (7):
\begin{align}\label{eq46}
\begin{bmatrix} A-I & B \\ HC & 0 \end{bmatrix} \begin{bmatrix} \bar{x}^{\ell^*} \\ \bar{u}^{\ell^*} \end{bmatrix}
=  \begin{bmatrix} -B_d \hat{d}^{\ell^*} \\ \bar{r}-HC_d \hat{d}^{\ell^*} \end{bmatrix}.
\end{align}
Let $\bar{x}_d^{\ell^*}$ denote the disturbance-augmented state at this point:
\begin{align}\label{eq47}
\bar{x}_d^{\ell^*}:=\begin{bmatrix} \bar{x}^{\ell^*}\\ \hat{d}^{\ell^*} \end{bmatrix}.
\end{align}
$\bar{x}_d^{\ell^*}$ is the equilibrium point of the proposed model-plant mismatch learning offset-free MPC. Therefore, we can analyze the closed-loop stability of the proposed scheme by examining the stability of the point $\bar{x}_d^{\ell^*}$.

Now, we can consider the optimal control problem $\mathrm{P_{\ell^*}}$ with $\bar{x}^{\ell^*}$ and $\bar{u}^{\ell^*}$.
\begin{align}
\mathrm{P_{\ell^*}}:\;J^*_{\ell^*}(x_d)=\underset{u_0,\cdots,u_{N-1}}{\mathrm{min}}\;& \phi_t^{\ell^*}(x_N)+\sum _{ i=0 }^{ N-1 }{\phi^{\ell^*}(x_i,u_i)} \nonumber\\
\mathrm{s.t.}\quad &x_{0}=x\nonumber\\
&x_{i+1}=Ax_{i}+Bu_{i}+B_d d \nonumber\\
&u_{i}\in{\mathcal{U}},\; x_{i+1}\in{\mathcal{X}},\; x_{N}\in {\mathcal{X}}_t \nonumber\\
&i=0,\dots ,N-1\nonumber
\end{align}
where $x_d:=[x^{\top},d^{\top}]^\top$ denotes the disturbance augmented state, $\phi^{\ell^*}(x_i,u_i):=|x_i-\bar{x}^{\ell^*}|^2_{Q_x}+|u_i-\bar{u}^{\ell^*}|^2_{Q_u}$ and $\phi_t^{\ell^*}(x_N):=|x_N-\bar{x}^{\ell^*}|^2_{Q_x^N}$ denote the stage cost and the terminal cost, respectively. The terminal constraint set $\mathcal{X}_t$ is chosen as a sublevel set of $\phi_t^{\ell^*}(\cdot)$:
\begin{align}\label{eq48}
\mathcal{X}_t=\{x\in\mathcal{X}\;|\; \phi_t^{\ell^*}(x)\leq \rho_t \}
\end{align}
for some $\rho_t >0$. Then, we specify the translated value function of P$_{\ell^*}$, $\bar{J}^*_{\ell^*}(x_d):={J}^*_{\ell^*}(x_d+\bar{x}_d^{\ell^*})$, as a Lyapunov function candidate.

\subsubsection{Perturbations in the combined system}

We define the perturbations in the combined system of model-plant mismatch learning offset-free MPC for further stability analysis. 

Let $e_{x_d^+}$ denote the prediction error on the evolved augmented state due to model-plant mismatch:
\begin{align}
&e_{x_d^+}:=\begin{bmatrix}f_p(S_x x_d,u) \\ S_d x_d \end{bmatrix}-f_m(x_d,u)\label{eq49} \\
&f_m(x_d,u):= \begin{bmatrix} A & B_d \\ 0 & I \end{bmatrix} x_d+ \begin{bmatrix} B \\ 0 \end{bmatrix} u.\label{eq50}
\end{align}
where $S_x:=[I_{n_x},\mathbf{0}_{n_x\times n_d}]$ and $S_d:=[\mathbf{0}_{n_d\times n_x},I_{n_d}]$. We consider the case in which the model-plant mismatch is not that severe such that $e_{x_d^+}$ is bounded:
\begin{align}\label{eq51}
|e_{x_d^+}|\leq c_{x_d^+}
\end{align}
where $c_{x_d^+}\in\mathbb{R}_{\geq 0}$.

Let $e_{\hat{x}_d}$ denote the estimation error between real augmented state $x_d$ and estimated augmented state $\hat{x}_d$:
\begin{align}\label{eq52}
e_{\hat{x}_d}:=x_d-\hat{x}_d.
\end{align}
Since we assumed that the pair (A,C) is observable in Section 2.1, the estimation error can be bounded as follows: $\delta_{\hat{x}_d}$ exists such that for all $||\mathbf{e}_{x_d^+}||\leq\delta_{\hat{x}_d}$ and $k\geq 0$, (53) holds from \textit{Appendix B}.
\begin{align}\label{eq53}
|e_{\hat{x}_d}(k)|\leq \beta_{\hat{x}_d}(|e_{\hat{x}_d}(0)|,k)+\sigma_{\hat{x}_d}(||\mathbf{e}_{x_d^+}||)
\end{align}
where $\beta_{\hat{x}_d}(\cdot)$ is a class $\mathcal{KL}$ function, $\sigma_{\hat{x}_d}(\cdot)$ is a class $\mathcal{K}$ function, and $\mathbf{e}_{x_d^+}$ denotes the sequence of $e_{x_d^+}$.

Unlike the system in \cite{23}, since the proposed offset-free MPC framework derives the target state and input for the optimal control problem based on the state and disturbance estimates as in (7), the proposed system has an additional perturbation in control law due to the target difference induced by the estimation error: 
\begin{align}\label{eq54}
e_{\kappa_p}:=\kappa_{p}-\kappa_{p}^{\ell^*}
\end{align}
where $\kappa_{p}^{\ell^*}$ and $\kappa_{p}^{\ell,s}$ denote the control laws from $\mathrm{P_{\ell^*}}$ and P, respectively. Then, $e_{\kappa_p}$ can be bounded as follows: there exists a class $\mathcal{K}$ function $\sigma_{\kappa_p^d}(\cdot)$ satisfying the inequality 
\begin{equation}\label{eq55}
|e_{\kappa_p}|\leq \sigma_{\kappa_p^d}(c_{\hat{d}^s})
\end{equation}
from \textit{Appendix C}.

Now, we can describe the estimate of successor state $\hat{x}_d^+$ through the proposed combined system with bounded perturbations in prediction, estimation, and control law:
\begin{align}\label{eq56}
&\hat{x}_d^+=f_m(\hat{x}_d+e_{\hat{x}_d},\kappa_p^{\ell^*}(\hat{x}_d)+e_{\kappa_p})+e_{x_d^+}-e_{\hat{x}_d}^+.
\end{align}

\subsubsection{Robust asymptotic stability of model-plant mismatch learning offset-free MPC}
In this section, we show the robust asymptotic stability of the equilibrium point $\bar{x}_d^{\ell^*}$ by establishing that $\bar{J}_{\ell^*}^*(\cdot)$ is an ISS Lyapunov function for the proposed combined system.

We can easily show that $\bar{J}^{*}_{\ell^*}$ satisfies the upper and lower bounds in (44) with \textbf{Proposition 5}:
\begin{align}\label{eq57}
\bar{\alpha}_1(|x_d-\bar{x}_d^{\ell^*}|)\leq \bar{J}_{\ell^*}^*(x_d-\bar{x}_d^{\ell^*}) \leq \bar{\alpha}_2(|x_d-\bar{x}_d^{\ell^*}|).
\end{align}\\
\textbf{Proposition 5}. Assume $V:\mathbb{R}^n\rightarrow \mathbb{R}$ is a continuous positive definite function defined on $\mathbb{R}^n$ and radially unbounded. Then, there exist class $\mathcal{K}_\infty$ functions $\alpha_1(\cdot)$ and $\alpha_2(\cdot)$ that satisfy the lower and upper bounds in (58).
\begin{align}\label{eq58}
\alpha_1(|\xi|)\leq V(\xi)\leq\alpha_2(|\xi|)
\end{align}
The proof is provided in \textit{Appendix C.4} of \cite*{27}.\\

Now, let $\tilde{x}_d^+$ denote the expected successor state with the control law $\kappa_p^{\ell^*}$ of the ideal problem $P_{\ell^*}$:
\begin{align}\label{eq59}
\tilde{x}_d^+:=f_m(\hat{x}_d,\kappa_p^{\ell^*}(\hat{x}_d)).
\end{align}
We first suggest a standard feasible solution $\tilde{\mathbf{u}}^+$ for $\tilde{x}_d^+$ and show that $\tilde{\mathbf{u}}^+$ is robustly feasible for $\hat{x}_d^+$ in (56) with all the perturbations. Then, we prove that $\bar{J}_{\ell^*}^*(\cdot)$ satisfies the inequality in (45) using $\tilde{\mathbf{u}}^+$ under the assumptions:\\
\quad\\
\textbf{Assumption 1}. The model-plant mismatch is not severe such that $\hat{d}^{\ell,s}$ is always in a compact set $\mathcal{D}\in\mathbb{R}^{n_d}$.\\
\quad\\
\textbf{Assumption 2}. With \textbf{Assumption 1}, let $\mathcal{X}^D_t:=\mathcal{X}_{t} \times \mathcal{D}$, $\phi_{d}^{\ell^*}(x_d,u):=\phi^{\ell^*}(S_x x_d,u)$, and $\phi_{d,t}^{\ell^*}(x_d):=\phi_t^{\ell^*}(S_x x_d)$. Then, for all $x_d\in\mathcal{X}^D_t$, there exists a local control law $\kappa_t: \mathcal{X}^D_t\rightarrow \mathcal{U}$ satisfying
\begin{align}
&f_m(x_d,\kappa_t(x_d))\in \mathcal{X}^D_t\label{eq60}\\
&\phi_{d,t}^{\ell^*}(f_m(x_d,\kappa_t(x_d)))\leq\phi_{d,t}^{\ell^*}(x_d)-\phi^{\ell^*}_d(x_d,\kappa_t(x_d)).\label{eq61}
\end{align}
\quad\\
\textbf{Assumption 3}. Let $\mathcal{X}^D:=\mathcal{X} \times \mathcal{D}$. There exists a class $\mathcal{K}_\infty$ function $\alpha_\phi(\cdot)$ satisfying the inequality in (62) for all $x_d\in\mathcal{X}^D$ and $u\in\mathcal{U}$.
\begin{align}\label{eq62}
\phi^{\ell^*}_d(x_d,u)\geq\alpha_{\phi}(|x_d-\bar{x}_d^{\ell^*}|).
\end{align}\\
\textbf{Proposition 6}. Let $e_c$ denote the perturbations in state transition for the combined disturbance estimator, target problem, and optimal control problem system in (56):
\begin{align}\label{eq63}
e_c:=(e_{\hat{x}_d},e_{\kappa_p},e_{x_d^+},e_{\hat{x}_d}^+).
\end{align} 
Then, for $\tilde{x}_d^+$ and $\hat{x}_d^+$, there exists a class $\mathcal{K}$ function $\sigma_{x^+}(\cdot)$ satisfying the inequality:
\begin{align}\label{eq64}
|\hat{x}_d^+ -\tilde{x}_d^+|\leq\sigma_{x^+}(|e_c|).
\end{align}
\textbf{Proof}. From (56) and (59), we have
\begin{align}\label{eq65}
|\hat{x}_d^+ -\tilde{x}_d^+|=&\;|f_m(\hat{x}_d+e_{\hat{x}_d},\kappa_p^{\ell^*}(\hat{x}_d)+e_{\kappa_p})+e_{x_d^+}-e_{\hat{x}_d}^+ \nonumber\\
&\quad-f_m(\hat{x}_d,\kappa_p^{\ell^*}(\hat{x}_d))| \nonumber\\
\leq&\;|f_m(\hat{x}_d+e_{\hat{x}_d},\kappa_p^{\ell^*}(\hat{x}_d)+e_{\kappa_p})-f_m(\hat{x}_d,\kappa_p^{\ell^*}(\hat{x}_d))| \nonumber\\
&\quad+|e_{x_d^+}|+|e_{\hat{x}_d}^+|.
\end{align}
Since $\mathcal{X}^D$ and $\mathcal{U}$ are compact sets in metric space and $f_m(x_d,u)$ is continuous, from \textbf{Proposition 4}, there exists a class $\mathcal{K}$ function $\sigma_{f_m}(\cdot)$ such that 
\begin{align}\label{eq66}
|f_m(\hat{x}_d+e_{\hat{x}_d},\kappa_p^{\ell^*}(\hat{x}_d)+e_{\kappa_p})-&f_m(\hat{x}_d,\kappa_p^{\ell^*}(\hat{x}_d))|\nonumber\\
&\leq\sigma_{f_m}(|(e_{\hat{x}_d},e_{\kappa_p})|).
\end{align}
By substituting (66) into (65) and applying (63), we obtain (67).
\begin{align}\label{eq67}
|\hat{x}_d^+ -\tilde{x}_d^+|&\leq\sigma_{f_m}(|(e_{\hat{x}_d},e_{\kappa_p})|)+|e_{x_d^+}|+|e_{\hat{x}_d}^+| \nonumber\\
&\leq\sigma_{f_m}(|e_c|)+|e_c|+|e_c| \nonumber\\
&\leq\sigma_{x^+}(|e_c|)
\end{align}
where $\sigma_{x^+}(z):=\sigma_{f_m}(z)+2z$.\qed\\
\quad\\
\textbf{Lemma 2}. Let $\tilde{\mathbf{u}}^{\ell^*}:=[\tilde{u}^{\ell^*\top}_0,\tilde{u}^{\ell^*\top}_1,\cdots,\tilde{u}^{\ell^*\top}_{N-1}]^\top$ denote the control sequence derived from P$_{\ell^*}$ for $\hat{x}_d$, and $\tilde{\mathbf{x}}_d^{\ell^*}:=[\tilde{x}_{d,1}^{\ell^*\top},\tilde{x}_{d,2}^{\ell^*\top},\cdots,\tilde{x}_{d,N}^{\ell^*\top}]^\top$ denote the resultant augmented state sequence where $\tilde{x}_{d,k}^{\ell^*}:=\eta(k,\hat{x}_d,\tilde{\mathbf{u}}^{\ell^*})$ and $\eta(k,x_d,\mathbf{u})$ denotes the open-loop solution of (50) for a given time step $k$ from $x_d$ with input sequence $\mathbf{u}$. We can construct a standard feasible solution for $\tilde{x}^+_d$ with a local control law $\kappa_t$ in \textbf{Assumption 2}:
\begin{align}\label{eq68}
\tilde{\mathbf{u}}^+:=[\tilde{u}^{\ell^*\top}_1,\tilde{u}^{\ell^*\top}_2,\cdots,\tilde{u}^{\ell^*\top}_{N-1},\kappa_t(\tilde{x}^{\ell^*}_{d,N})^\top]^\top.
\end{align}
Then, $\tilde{\mathbf{u}}^+$ is also robustly feasible for $\hat{x}_d^+$ when the perturbation $|e_c|$ is sufficiently small.\\
\quad\\ 
\textbf{Proof}. Since $\tilde{x}_{d,N}^{\ell^*}\in\mathcal{X}^D_t$, (69) holds from (61) and (62) in \textbf{Assumptions 2} and \textbf{3}.
\begin{align}\label{eq69}
\phi_{d,t}^{\ell^*}(f_m(\tilde{x}_{d,N}^{\ell^*},\kappa_t(\tilde{x}_{d,N}^{\ell^*})))\leq\phi_{d,t}^{\ell^*}(\tilde{x}_{d,N}^{\ell^*})-\alpha_\phi(|\tilde{x}_{d,N}^{\ell^*}-\bar{x}_d^{\ell^*}|).
\end{align}
Let $\hat{x}^+_{d,k}:=\eta(k,\hat{x}^+_d,\tilde{\mathbf{u}}^+)$ and $\tilde{x}_{d,k}^+:=\eta(k,\tilde{x}_d^+,\tilde{\mathbf{u}}^+)$. Since $\tilde{x}^+_{d,N-1}=\tilde{x}_{d,N}^{\ell^*}$, (70) holds.
\begin{align}\label{eq70}
\tilde{x}^+_{d,N}=f_m(\tilde{x}_{d,N}^{\ell^*},\kappa_t(\tilde{x}_{d,N}^{\ell^*})).
\end{align}
Then, substituting (70) into (69) yields
\begin{align}\label{eq71}
\phi_{d,t}^{\ell^*}(\tilde{x}^+_{d,N})\leq\phi_{d,t}^{\ell^*}(\tilde{x}_{d,N}^{\ell^*})-\alpha_\phi(|\tilde{x}_{d,N}^{\ell^*}-\bar{x}_d^{\ell^*}|).
\end{align}

Since $\phi_{d,t}^{\ell^*}(\cdot)$ is a continuous function, from \textbf{Proposition 4}, there exists a class $\mathcal{K}$ function $\sigma_t(\cdot)$ such that
\begin{align}\label{eq72}
|\phi_{d,t}^{\ell^*}(x_d^1)-\phi_{d,t}^{\ell^*}(x_d^2)|\leq\sigma_t(|x_d^1-x_d^2|).
\end{align}
Since $\phi_{d,t}^{\ell^*}(\bar{x}_d^{\ell^*})=0$, substituting $\tilde{x}^{\ell^*}_{d,N}$ and $\bar{x}_d^{\ell^*}$ into (72) yields
\begin{align}\label{eq73}
\phi_{d,t}^{\ell^*}(\tilde{x}^{\ell^*}_{d,N})\leq\sigma_t(|\tilde{x}^{\ell^*}_{d,N}-\bar{x}_d^{\ell^*}|)
\end{align}
for all $\tilde{x}^{\ell^*}_{d,N}\in\mathcal{X}^D_t$. Substituting (73) into (71) yields
\begin{align}\label{eq74}
&\phi_{d,t}^{\ell^*}(\tilde{x}^+_{d,N})\leq \alpha_{t,\phi}(|\tilde{x}_{d,N}^{\ell^*}-\bar{x}_d^{\ell^*}|)\\
&\alpha_{t,\phi}(z):=\sigma_{t}(z)-\alpha_{\phi}(z)\nonumber.
\end{align}

Now, substituting $\hat{x}_{d,N}^+$ and $\tilde{x}_{d,N}^+$ into (72) yields
\begin{align}\label{eq75}
|\phi_{d,t}^{\ell^*}(\hat{x}_{d,N}^+)-\phi_{d,t}^{\ell^*}(\tilde{x}_{d,N}^+)|\leq\sigma_t(|\hat{x}_{d,N}^+-\tilde{x}_{d,N}^+|).
\end{align}
Since $\sigma_t(\cdot)$ is a class $\mathcal{K}$ function, (76) holds for both cases where $\phi_{d,t}^{\ell^*}(\hat{x}_{d,N}^+)-\phi_{d,t}^{\ell^*}(\tilde{x}_{d,N}^+)$ is either positive or negative.
\begin{align}\label{eq76}
\phi_{d,t}^{\ell^*}(\hat{x}_{d,N}^+)\leq\phi_{d,t}^{\ell^*}(\tilde{x}_{d,N}^+)+\sigma_t(|\hat{x}_{d,N}^+-\tilde{x}_{d,N}^+|).
\end{align}
Since $\hat{x}^+_{d,k}=\eta(k,\hat{x}^+_d,\tilde{\mathbf{u}}^+)$, $\tilde{x}_{d,k}^+=\eta(k,\tilde{x}_d^+,\tilde{\mathbf{u}}^+)$, and $\eta(\cdot)$ is a continuous function, there exists a class $\mathcal{K}$ function $\sigma_\eta(\cdot)$ such that
\begin{align}\label{eq77}
|\hat{x}_{d,N}^+-\tilde{x}_{d,N}^+| \leq \sigma_\eta(|\hat{x}_{d}^+-\tilde{x}_{d}^+|).
\end{align}
Substituting (64) in \textbf{Proposition 6} and (77) into (76) yields
\begin{align}\label{eq78}
&\phi_{d,t}^{\ell^*}(\hat{x}_{d,N}^+)\leq\phi_{d,t}^{\ell^*}(\tilde{x}_{d,N}^+)+\sigma_{t,\eta,x^+}(|e_c|) \\
&\sigma_{t,\eta,x^+}(|e_c|):=\sigma_t(\sigma_\eta(\sigma_{x^+}(|e_c|))). \nonumber
\end{align}

Now, substituting (74) into (78) yields
\begin{align}\label{eq79}
\phi_{d,t}^{\ell^*}(\hat{x}_{d,N}^+)\leq\alpha_{t,\phi}(|\tilde{x}_{d,N}^{\ell^*}-\bar{x}_d^{\ell^*}|)+\sigma_{t,\eta,x^+}(|e_c|). 
\end{align}
Since $\mathcal{X}^D_t=\{x_d|\;\phi_{d,t}^{\ell^*}(x_d)\leq\rho_t\}$ from (48), when $|e_c|$ is sufficiently small as (80) holds, $\hat{x}_{d,N}^+\in\mathcal{X}^D_t$.
\begin{align}\label{eq80}
|e_c|\leq\sigma_{t,\eta,x^+}^{-1}(\rho_t-\alpha_{t,\phi}(|\tilde{x}_{d,N}^{\ell^*}-\bar{x}_d^{\ell^*}|)).
\end{align}

\qed\\
\quad\\
\textbf{Lemma 3}. Assume $\hat{x}_d\in\mathcal{X}^D:=\{ x_d|\;\bar{J}^*_{\ell^*}(x_d-\bar{x}_d^{\ell^*})\leq\rho_J\}$. Then, (81) holds with a class $\mathcal{K}_\infty$ function $\alpha(\cdot)$ and a class $\mathcal{K}$ function $\sigma(\cdot)$, and $\mathcal{X}^D$ is robustly positive invariant when the perturbation $|e_c|$ is sufficiently small.
\begin{align}\label{eq81}
\bar{J}_{\ell^*}^*(\hat{x}_d^+-\bar{x}_d^{\ell^*})\leq \bar{J}_{\ell^*}^*(\hat{x}_d-\bar{x}_d^{\ell^*})-\alpha(|\hat{x}_d-\bar{x}_d^{\ell^*}|)+\sigma(||\mathbf{e}_c||)
\end{align}
where $\mathbf{e}_c$ is the sequence of the perturbation $e_c$ in (63).\\
\quad\\
\textbf{Proof}. Let $x_{d,k}:=\eta(k,x_d,\mathbf{u})$ and $J_{\ell^*}(x_d,\mathbf{u}):=\phi_{d,t}^{\ell^*}(x_{d,N})+\sum _{ i=0 }^{ N-1 }{\phi_d^{\ell^*}(x_{d,i},u_i)}$ denote the resultant cost sum from $x_d$ with the solution sequence $\mathbf{u}$. Then, we can derive (82).
\begin{align}\label{eq82}
&J_{\ell^*}(\tilde{x}_d^+,\mathbf{u}^+)-\bar{J}_{\ell^*}^*(\hat{x}_d-\bar{x}_d^{\ell^*})+\phi^{\ell^*}_{d}(\hat{x}_d,\kappa_p^{\ell^*}(\hat{x}_d)) \nonumber\\
&=\phi_d^{\ell^*}(\tilde{x}_{d,N}^{\ell^*},\kappa_t(\tilde{x}_{d,N}^{\ell^*}))+\phi^{\ell^*}_{d,t}(\tilde{x}^+_{d,N})-\phi^{\ell^*}_{d,t}(\tilde{x}_{d,N}^{\ell^*})
\end{align}
From  (61) in \textbf{Assumption 2}, the right-hand side of (82) yields
\begin{align}\label{eq83}
\phi_d^{\ell^*}(\tilde{x}_{d,N}^{\ell^*},\kappa_t(\tilde{x}_{d,N}^{\ell^*}))+\phi^{\ell^*}_{d,t}(\tilde{x}^+_{d,N})-\phi^{\ell^*}_{d,t}(\tilde{x}_{d,N}^{\ell^*})\leq 0.
\end{align}
Substituting (83) into (82) and rearranging yields
\begin{align}\label{eq84}
J_{\ell^*}(\tilde{x}_d^+,\mathbf{u}^+)\leq \bar{J}_{\ell^*}^*(\hat{x}_d-\bar{x}_d^{\ell^*})-\phi^{\ell^*}_{d}(\hat{x}_d,\kappa_p^{\ell^*}(\hat{x}_d)).
\end{align}
Then, by substituting (62) in \textbf{Assumption 3} into (84), we can derive
\begin{align}\label{eq85}
J_{\ell^*}(\tilde{x}_d^+,\mathbf{u}^+)\leq \bar{J}_{\ell^*}^*(\hat{x}_d-\bar{x}_d^{\ell^*})-\alpha_\phi(|\hat{x}_d-\bar{x}_d^{\ell^*}|).
\end{align}

Now, since $J_{\ell^*}(x_d,u)$ is continuous, there exists a class $\mathcal{K}$ function $\sigma_J(\cdot)$ from \textbf{Proposition 4}:
\begin{align}\label{eq86}
|J_{\ell^*}(\hat{x}_d^+,\mathbf{u}^+)-J_{\ell^*}(\tilde{x}_d^+,\mathbf{u}^+)|\leq \sigma_J(|\hat{x}_d^+-\tilde{x}_d^+|).
\end{align}
We can drop the absolute value of (86) in the similar manner to obtain (76) from (75):
\begin{align}\label{eq87}
J_{\ell^*}(\hat{x}_d^+,\mathbf{u}^+)\leq J_{\ell^*}(\tilde{x}_d^+,\mathbf{u}^+)+\sigma_J(|\hat{x}_d^+-\tilde{x}_d^+|).
\end{align}
Substituting (64) in \textbf{Proposition 6} and (85) into (87) yields
\begin{align}\label{eq88}
J_{\ell^*}(\hat{x}_d^+,\mathbf{u}^+)\leq \bar{J}_{\ell^*}^*(\hat{x}_d-\bar{x}_d^{\ell^*})-\alpha_\phi(|\hat{x}_d-\bar{x}_d^{\ell^*}|)+\sigma_{J,x^+}(|e_c|)
\end{align}
where $\sigma_{J,x^+}(|e_c|):=\sigma_{J}(\sigma_{x^+}(|e_c|))$ is a class $\mathcal{K}$ function. Since $\bar{J}_{\ell^*}^*(\hat{x}_d^+-\bar{x}_d^{\ell^*})\leq J_{\ell^*}(\hat{x}_d^+,\mathbf{u}^+)$ and $|e_c|\leq||\mathbf{e}_c||$, (89) holds.
\begin{align}\label{eq89}
\bar{J}_{\ell^*}^*(\hat{x}_d^+-\bar{x}_d^{\ell^*})\leq &\bar{J}_{\ell^*}^*(\hat{x}_d-\bar{x}_d^{\ell^*}) \\
&-\alpha_\phi(|\hat{x}_d-\bar{x}_d^{\ell^*}|)+\sigma_{J,x^+}(||\mathbf{e}_c||).\nonumber
\end{align}
Therefore, we can see that (81) holds.

Additionally, when $|e_c|$ is sufficiently small as (90) holds, $\hat{x}_d\in\mathcal{X}^D$ implies $\hat{x}_d^+\in\mathcal{X}^D$.
\begin{align}\label{eq90}
|e_c|\leq\sigma_{J,x^+}^{-1}(\rho_J-\bar{J}_{\ell^*}^*(\hat{x}_d-\bar{x}_d^{\ell^*})+\alpha_\phi(|\hat{x}_d-\bar{x}_d^{\ell^*}|)).
\end{align}\qed\\
\quad\\
\textbf{Theorem 4}. From \textbf{Lemmas 2} and \textbf{3}, we can establish that the translated value function $\bar{J}^*_{\ell^*}(\cdot)$ of the ideal optimal control problem P$_{\ell^*}$ with the ideal target pair $(\bar{x}^{\ell^*},\bar{u}^{\ell^*})$ is an ISS Lyapunov function in the robust invariant set $\mathcal{X}^D$ at the combined system of the proposed model-plant mismatch learning offset-free MPC. Finally, by \textbf{Proposition 3}, we can see that the equilibrium point $\bar{x}_d^{\ell^*}$ of the proposed combined system is robustly asymptotically stable.

\section{Numerical example}\label{sec4}

In this section, we present three numerical simulation results to demonstrate the closed-loop performance of the proposed model-plant mismatch learning offset-free MPC. The first case shows the closed-loop simulation results when a setpoint of a controlled variable changes, the second case shows the results of the proposed scheme with data window when the plant characteristics change during operation, and the third case shows the results when setpoints for multiple variables change with time. 

We consider the continuous stirred-tank reactor (CSTR) presented in \cite{24}. The first-order reaction, $A\rightarrow B$ takes place in the liquid phase, and the reactor temperature is controlled with the external cooling jacket. 

The control objective is to track the reference trajectories of the outlet concentration of the reactant ($c$) while regulating the reactor temperature ($T$) as a fixed value by directly manipulating the jacket temperature ($T_c$) and the outlet flow rate ($F$). The following equations describe the dynamics
of the reactor.
\begin{align}\label{eq91}
\begin{cases} \dfrac{dc}{dt}=\dfrac{F_0 (c_0-c)}{\pi r^2 h}-k_0\; {\mathrm{exp}}(-\dfrac{E}{RT})c \\[10pt]
\dfrac{dT}{dt}=\dfrac{F_0 (T_0 -T)}{\pi r^2 h}-\dfrac{\Delta H}{\rho C_p}k_0\; {\mathrm{exp}}(-\dfrac{E}{RT})c \\ \qquad\quad +\dfrac{2U}{r\rho C_p}(T_c -T) \\[10pt]
\dfrac{dh}{dt}=\dfrac{F_0-F}{\pi r^2}. \end{cases} 
\end{align}
The parameters are given in \textbf{Table 1}.

\begin{table}[t!]
\begin{center} 
\caption{Parameters applied to the simulation of a CSTR.}
\begin{tabular}{l|l|l}
\hline
Parameter & Value & Unit \\ \hline
$F_0$ & 0.1 & $\mathrm{m^3 /min}$ \\
$T_0$ & 350 & $\mathrm{K}$\\
$c_0$ & 1 & $\mathrm{kmol/m^3}$\\
$r$ & 0.219 & $\mathrm{m}$\\
$k_0$ & 7.2$\times 10^{10}$ & $\mathrm{min^{-1}}$\\
$E/R$ & 8750 & $\mathrm{K}$\\
$U$ & 54.94 & $\mathrm{kJ/min\cdot m^2 \cdot K}$\\
$\rho$ & 1000 & $\mathrm{kg/m^3}$\\
$C_p$ & 0.239 & $\mathrm{kJ/kg\cdot K}$\\
$\Delta H$ & -5$\times 10^4$ & $\mathrm{kJ/kmol}$\\ \hline 
\end{tabular}
\end{center}
\end{table}

The linear model is derived at the steady state:
\begin{align}
&c^s =0.878 \;{\mathrm{kmol/m^3}},\quad T^s =324.5\;{\mathrm{K}},\quad h^s=0.659\;{\mathrm{m}} \nonumber\\
&T_c^s =300\;{\mathrm{K}},\quad F^s=0.1\;{\mathrm{m^3/min}} \nonumber
\end{align}

The discretized linear model in (92) with sampling instant of 1 min is used for MPC.
\begin{align}\label{eq92}
\begin{cases} x(k+1)=A x(k)+B u(k) \\ y(k)=C x (k) \\ z(k)=Hy(k) \end{cases} 
\end{align}
\vspace{-0.2cm}
\begin{align}
&A=\begin{bmatrix} 0.2681 & -0.00338 & -0.00728 \\ 9.703 & 0.3279 & -25.44 \\ 0 & 0 & 1 \end{bmatrix}, \nonumber\\
&B=\begin{bmatrix} -0.00537 & 0.1655 \\ 1.297 & 97.91 \\ 0 & -6.637 \end{bmatrix}, \nonumber\\
&C=\begin{bmatrix} 1 & 0 & 0 \\ 0 & 1 & 0 \\ 0 & 0 & 1 \end{bmatrix},\; 
H=\begin{bmatrix} 1 & 0 & 0 \\ 0 & 1 & 0 \end{bmatrix}. \nonumber
\end{align}

The following operational constraints are applied to the system:
\begin{align}
&0.83\leq c\leq 0.92,\; 320\leq T\leq 330,\; 0.4 \leq h\leq 1.2, \nonumber\\
&295\leq T_c \leq 310,\;\mathrm{and}\; 0.07 \leq F \leq 0.13 \nonumber
\end{align}

The prediction horizon and weights in the optimal control problem are $N=10$, $Q_x = diag \{ 50; 0.001; 1 \rbrace$, $Q_x^N = 10Q_x \rbrace$ and $Q_u = diag \lbrace 0.01; 0.01 \rbrace$.

For the disturbance estimator, the following disturbance model and gain matrices are applied:
\begin{align}
&B_d=\begin{bmatrix} 1 & 0 \\ 0 & 1 \\ 0 & 0 \end{bmatrix},\; C_d={\mathbf{0}}_{n_y \times n_d}, \nonumber\\
&L_x=\begin{bmatrix} 0.6141 & -0.0034 & -0.0071 \\ 9.7056 & 0.7346 & -25.4413 \\ 0 & 0 & 0.2800 \end{bmatrix}, \nonumber\\
&L_d= \begin{bmatrix} 0.3420 & 0 & 0 \\ 0 & 0.4026 & 0 \end{bmatrix}. \nonumber
\end{align}

We also assume that the outlet concentration is $3\%$ higher than the average concentration in the reactor, $F \rightarrow 1.03\:F$, to apply additional error and increase the model-plant mismatch.

\begin{figure}[t!]
\begin{center}
\includegraphics[width=7cm]{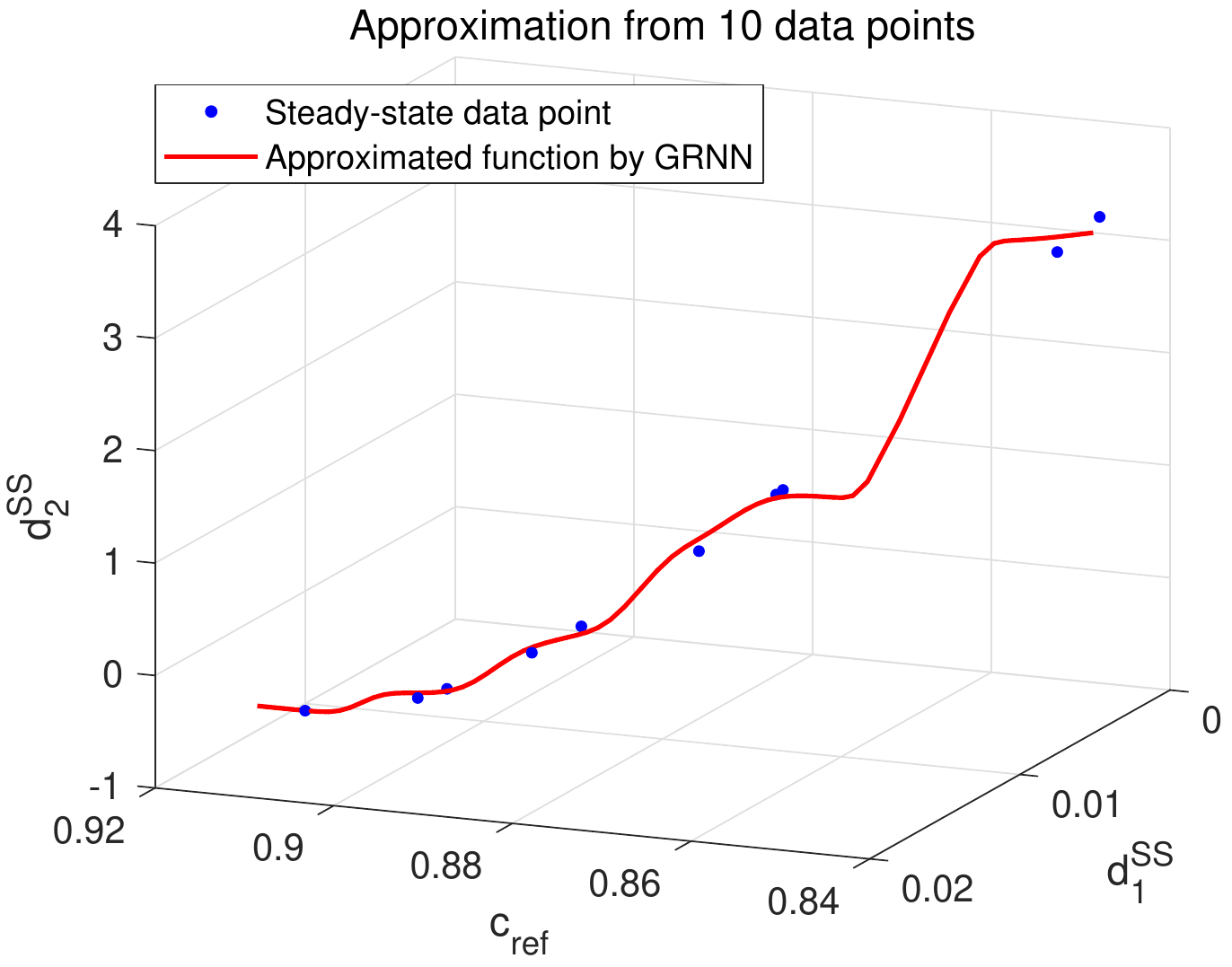}
\includegraphics[width=7cm]{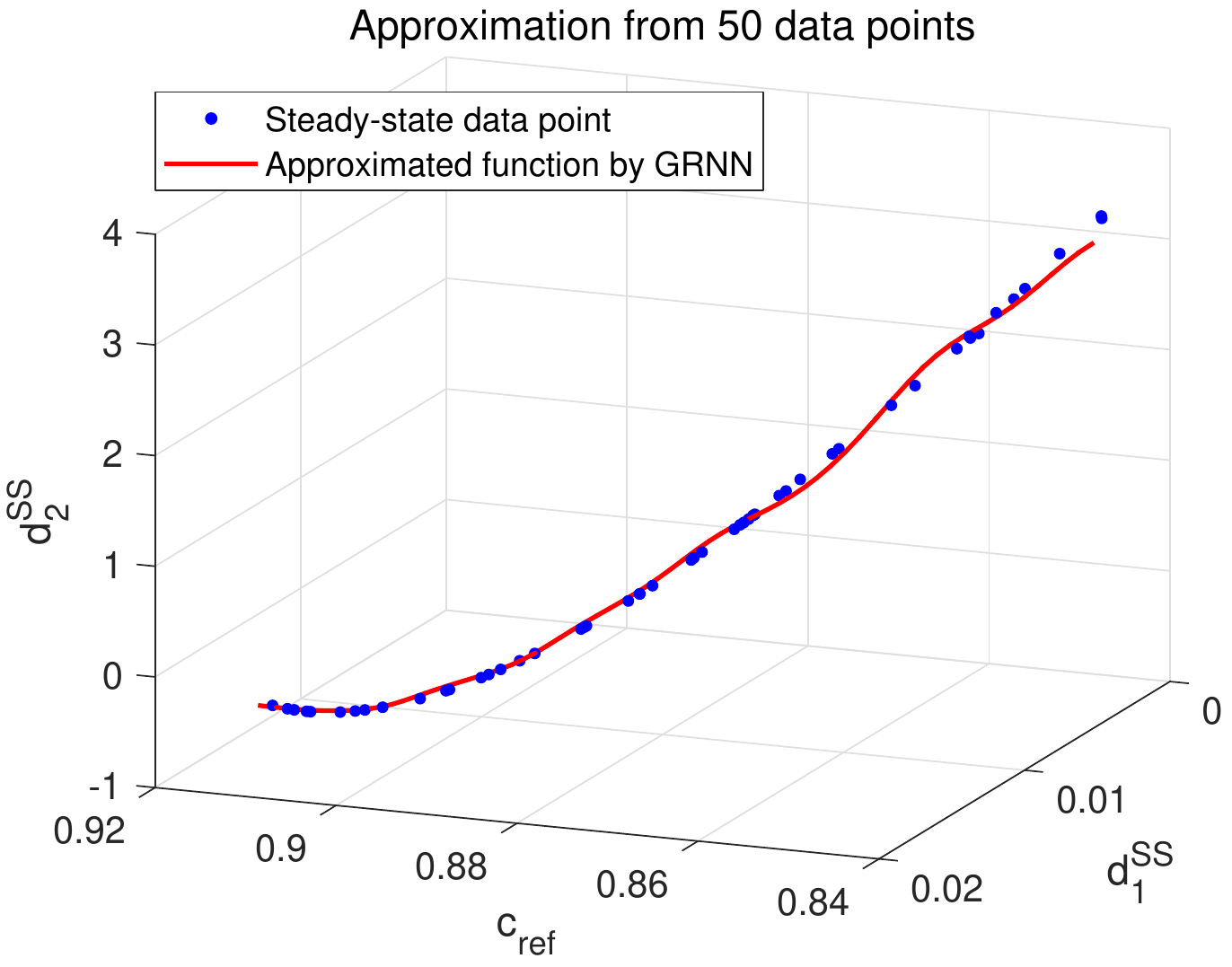}
\caption{Approximated model-plant mismatch by GRNN from 10 and 50 steady-state data points.}
\end{center}
\end{figure}

\subsection{System with setpoint change}

Fig.~1 illustrates the results of function approximation for steady-state model-plant mismatch $f_d$ with GRNN from 10 and 50 steady-state data points. We can see the estimated disturbance data and setpoint shows a certain relation that indicates the steady-state model-plant mismatch for each setpoint between (91) and (92). The approximated function by GRNN follows the trend of the data well with sufficient data.

\begin{figure}[t!]
\begin{center}
\includegraphics[width=9cm]{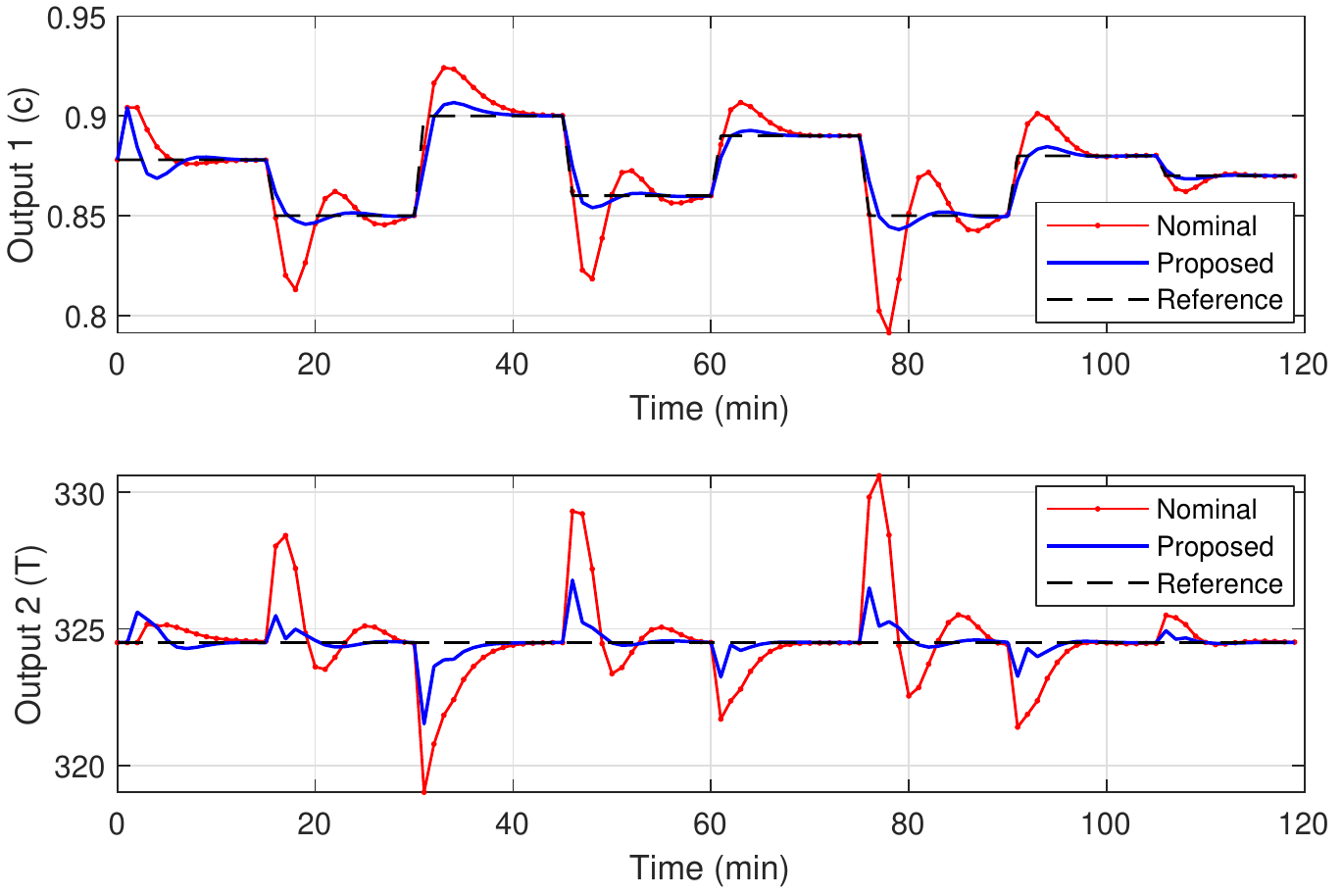}
\caption{Reference tracking results of offset-free MPC schemes with the changed setpoint for $c$ and the fixed setpoint for $T$.}
\end{center}
\end{figure}

\begin{figure}[t!]
\begin{center}
\includegraphics[width=9cm]{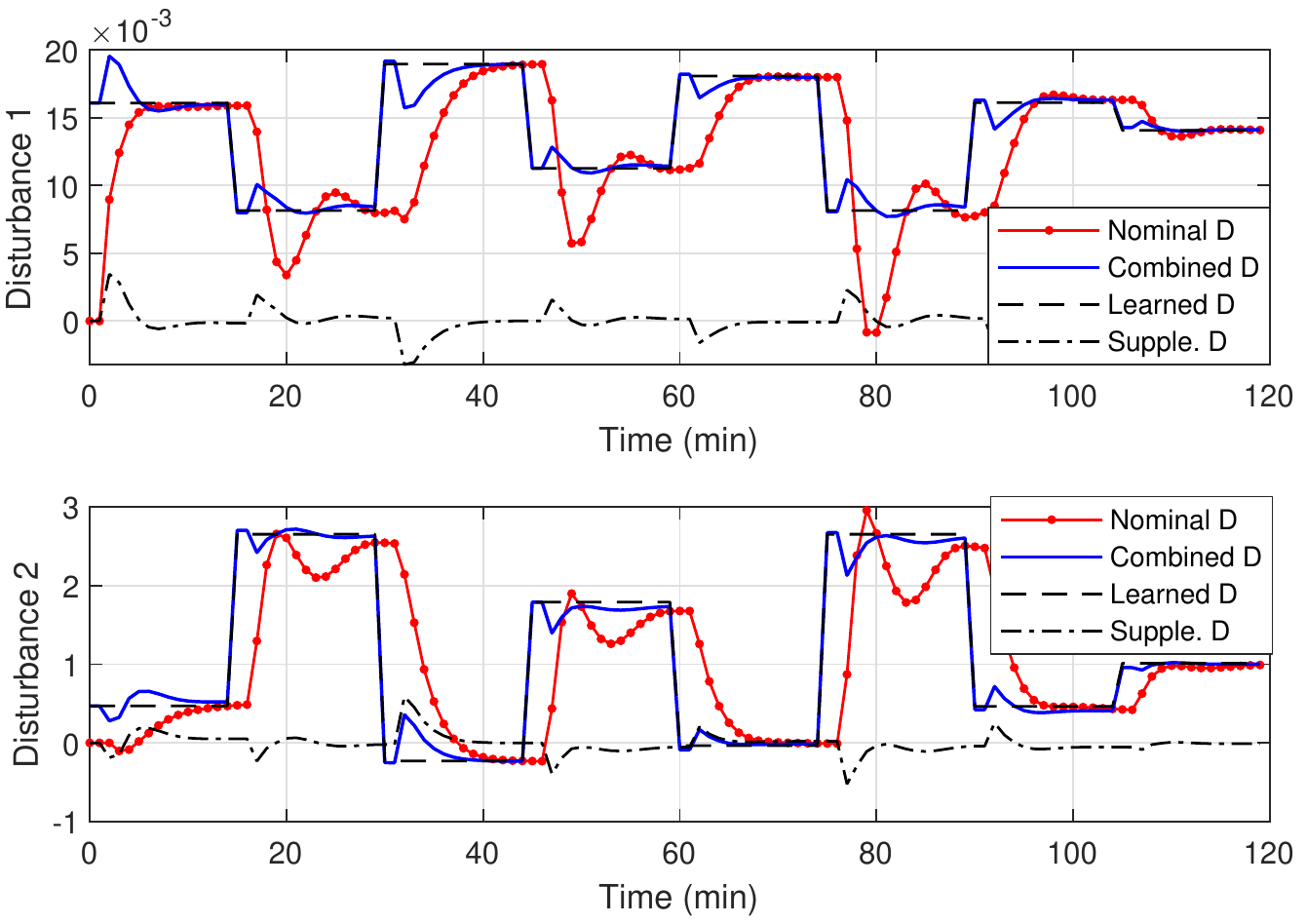}
\caption{Estimated disturbance trajectories of each offset-free MPC scheme with the changed setpoint for $c$ and the fixed setpoint for $T$.}
\end{center}
\end{figure}

\begin{figure}[t!]
\begin{center}
\includegraphics[width=6.8cm]{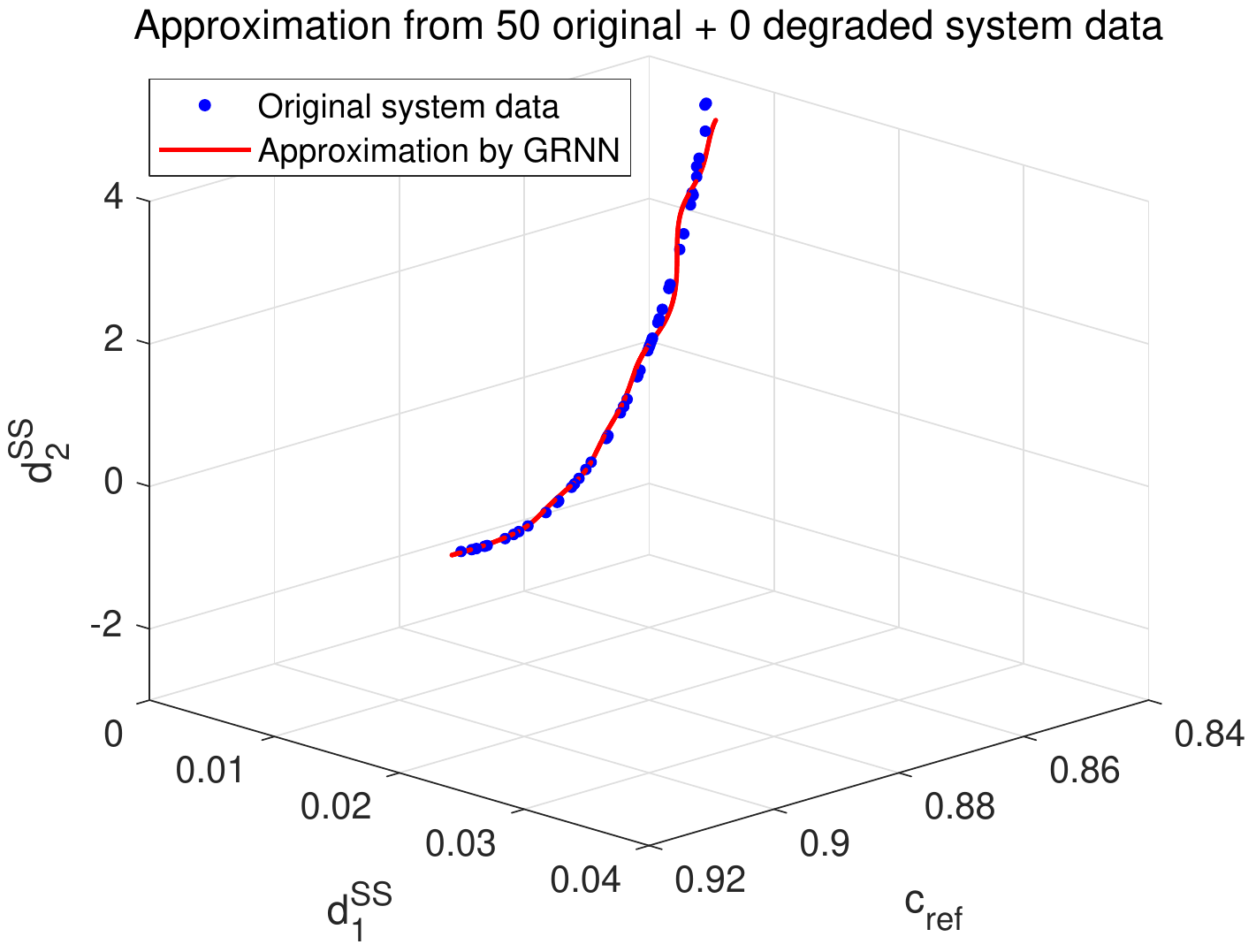}\vspace{0.15cm}\\
\includegraphics[width=6.8cm]{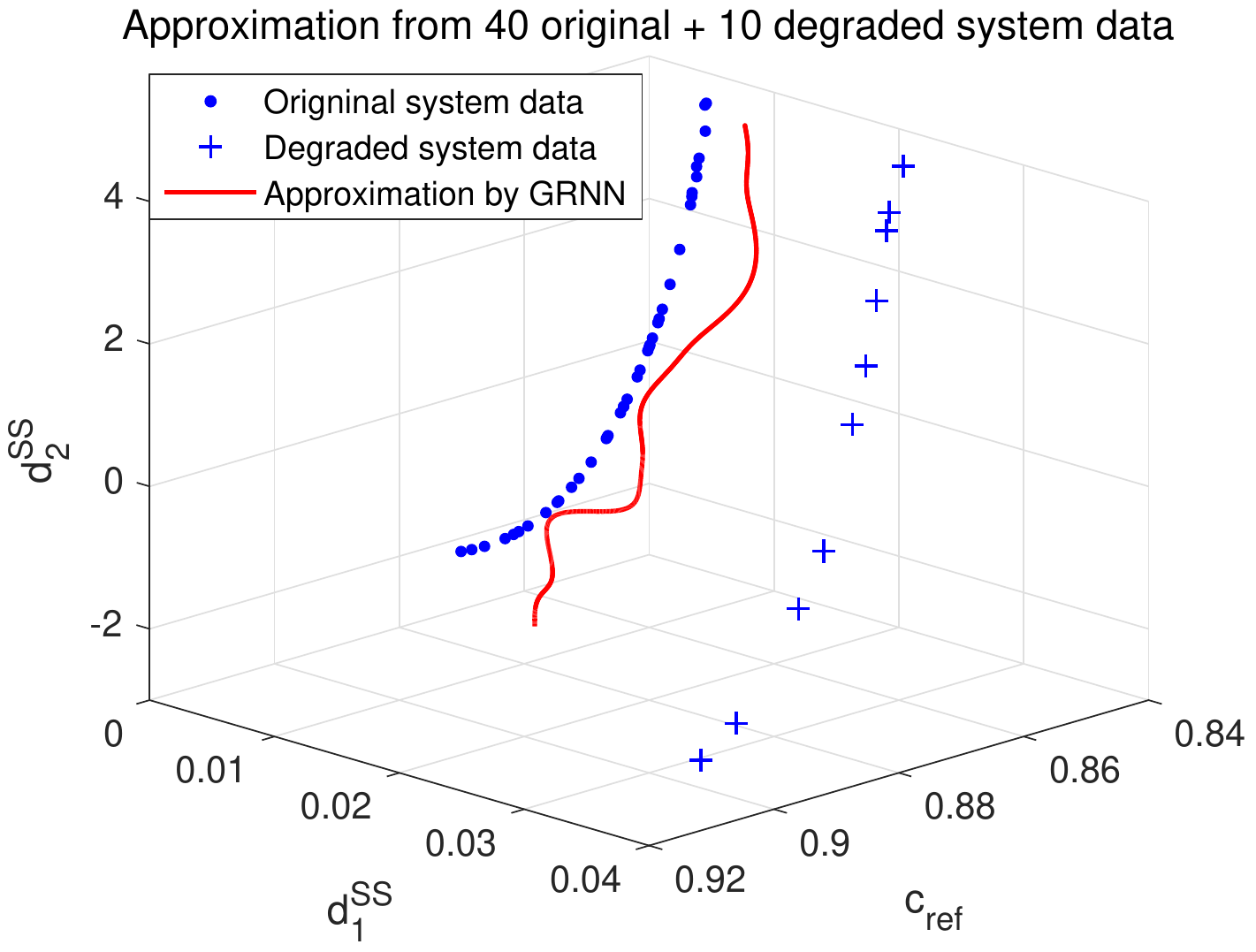}\vspace{0.15cm}\\
\includegraphics[width=6.8cm]{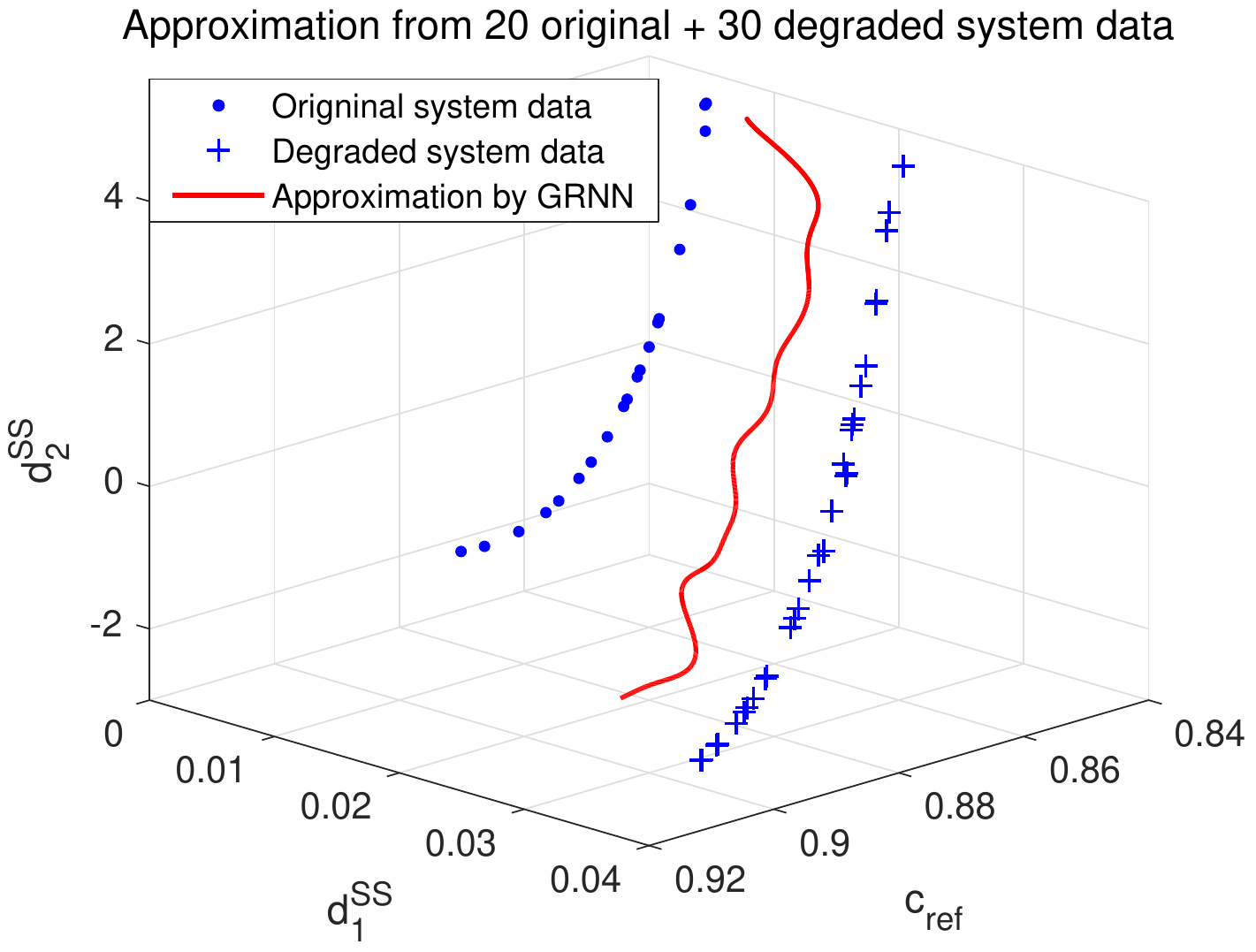}\vspace{0.15cm}\\
\includegraphics[width=6.8cm]{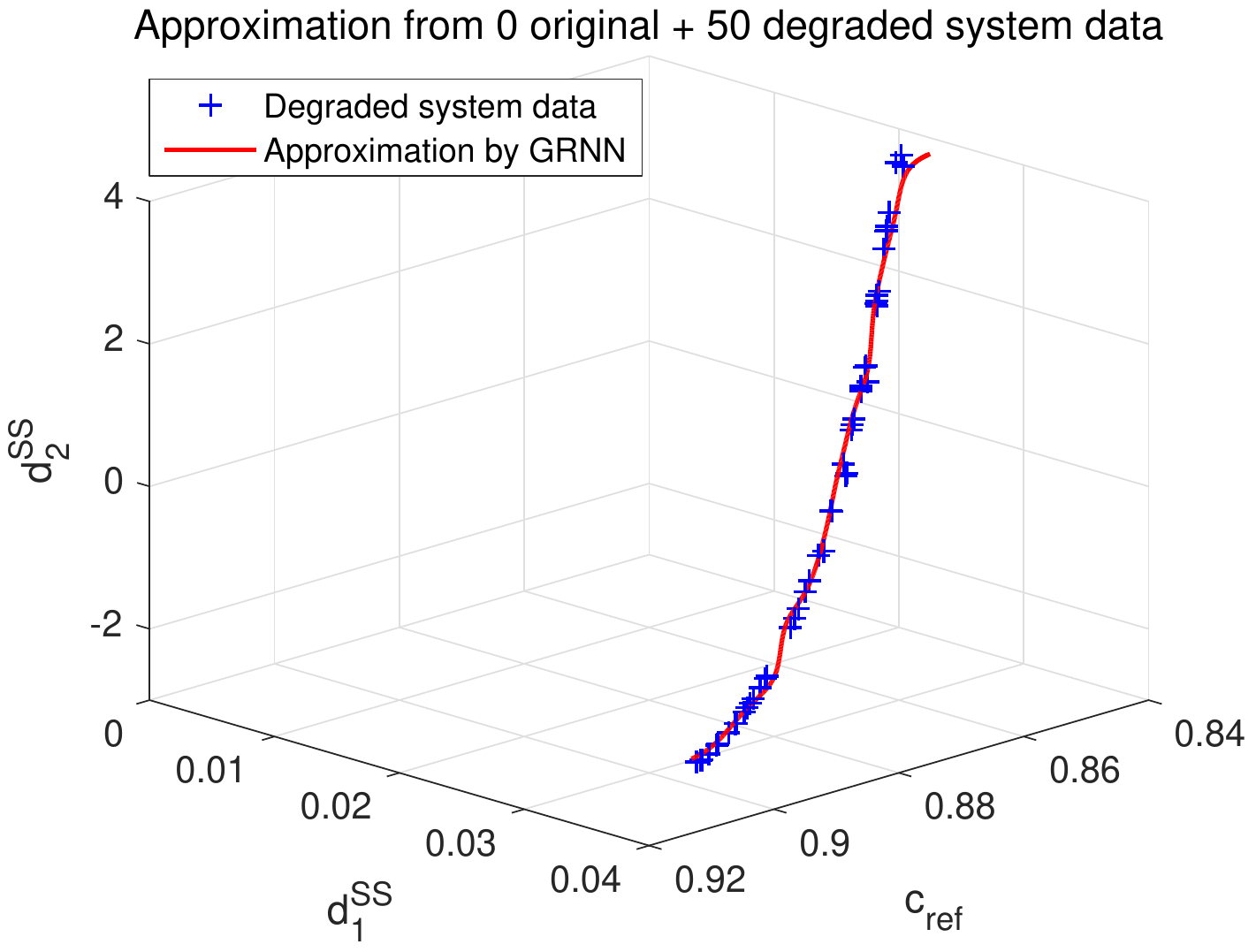}
\caption{Transformation of the approximated model-plant mismatch under system characteristic change.}
\end{center}
\end{figure}

Fig.~2 shows the reference tracking results of the nominal offset-free MPC and the proposed model-plant mismatch learning offset-free MPC which applies a learned disturbance value approximated by GRNN from 50 steady-state data points in Fig.~1 when the setpoint for $c$ changes along $[0.84,0.91]$ for every 15 min while the setpoint for $T$ is fixed. Although all the schemes achieve offset-free tracking, the nominal scheme shows considerable deviation from the setpoint because the nominal method gradually estimates the model-plant mismatch for each setpoint only from the error between prediction and measurement. However, the proposed scheme tracks the reference trajectory with slight deviation from the setpoint at the transient state and shows much better tracking performance compared to the standard method because the proposed scheme learns the intrinsic model-plant mismatch from the data and utilizes the learned information in the estimator and finite-horizon optimal control problem.

From Fig.~3 which describes the estimated disturbance values of each scheme, we can examine how the proposed scheme improves the controller performance in more detail. `Nominal D' and `Combined D' imply the estimated disturbance from the nominal offset-free MPC and the proposed model-plant mismatch learning offset-free MPC, respectively. We can see that the proposed scheme shows considerably superior performance in disturbance estimation. `Combined D' is the sum of learned disturbance value with GRNN, `Learned D', and the supplementary disturbance value, which is estimated from the estimator in (27), `Supple. D'. The learned disturbance value acts as a warm-start signal so that the supplementary disturbance value changes only in a much smaller range than the fully estimated disturbance value of the nominal scheme. Thus, the proposed scheme effectively improves the model-plant mismatch compensating performance.

\subsection{System with characteristics change}

In practice, system characteristics can change during operation. In this section, we apply a data window, which keeps the most recent data and discards old data, to model-plant mismatch learning process and examine the efficacy of the proposed scheme when system characteristics change. 

Fig.~4 illustrates the result of approximation for the steady-state model-plant mismatch with the data-window strategy when the degradation of the reaction rate occurs ($k_0=7.2\times 10^{10}\mathrm{min^{-1}}\rightarrow k_0=6.2\times 10^{10}\mathrm{min^{-1}}$). The approximated model-plant mismatch function $\hat{f}_d(\cdot)$ is gradually transformed from that of the original system to that of the degraded system as the sampled data from the original system is replaced with the data from the degraded system. From this, we can see that the proposed scheme with the data-window strategy can effectively approximate the model-plant mismatch under system characteristic change.

\begin{figure}[t!]
\begin{center}
\includegraphics[width=9cm]{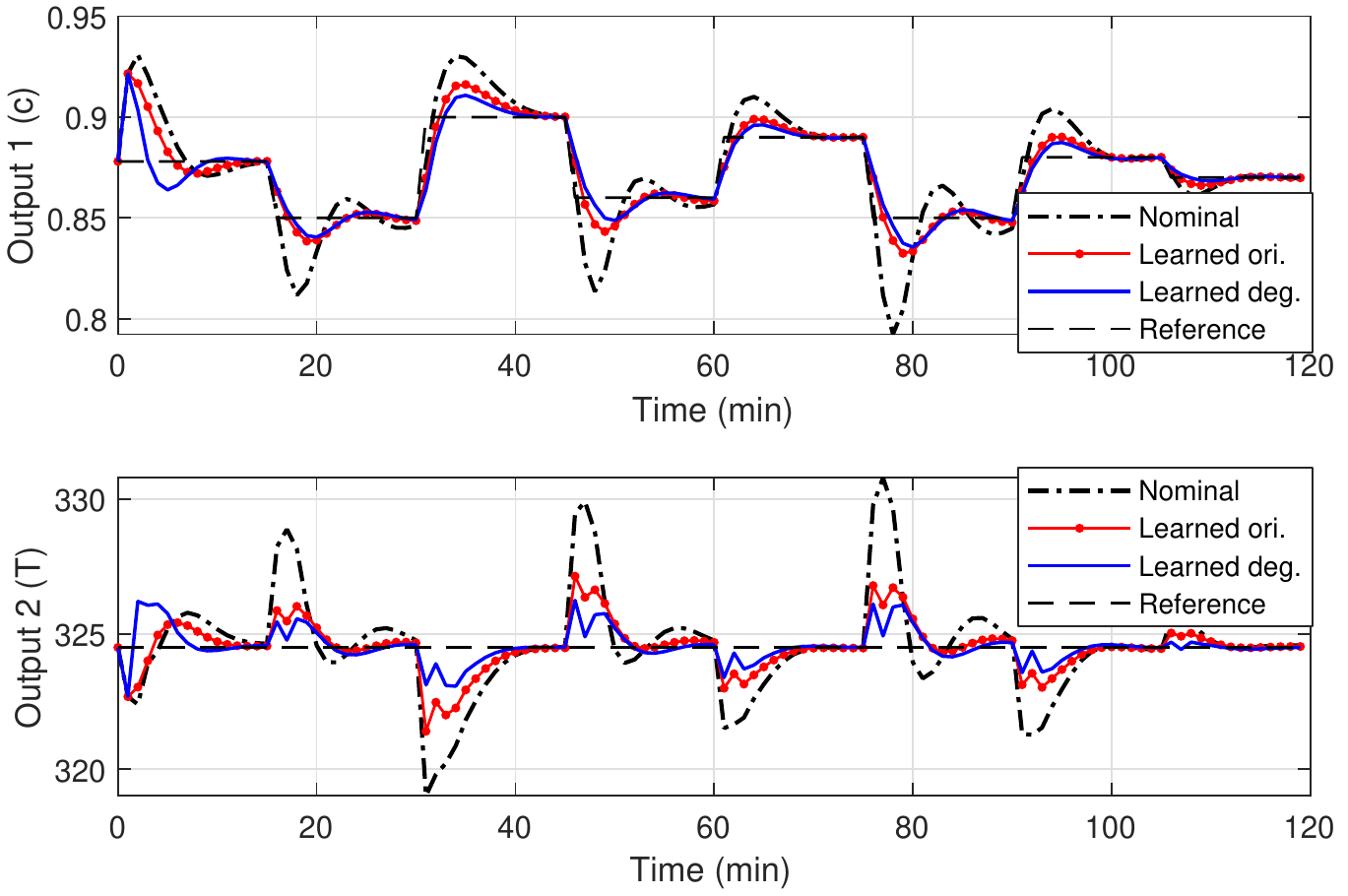}
\caption{Reference tracking results of offset-free MPC schemes with the changed setpoint for $c$ and the fixed setpoint for $T$ in the transformed system.}
\end{center}
\end{figure}

Fig.~5 describes the reference tracking result of nominal and proposed offset-free MPC schemes where `Learned ori.' and `Learned deg.' denote the closed-loop results with the learned model-plant mismatch from the 50 original system data points and the 50 degraded system data points in Fig.~4, respectively. We can see that the `Learned deg.' shows proper reference tracking performance with small deviation from the setpoint in the transient state; thus, the proposed scheme with the data window strategy is efficient under system characteristics change. Furthermore, in the case of the trajectories from `Learned ori.', although the deviation in the transient state is larger than that of `Learned deg.', they also show proper reference tracking performance due to the robustness of the proposed scheme described in Section~\ref{sec3}.3.

Fig.~6 shows the estimated disturbance values of each scheme. The learned disturbance value from the degraded system data in Fig.~6(b) matches the steady-state disturbance well. However, the learned disturbance value from the original system data in Fig.~6(a) shows deviations from the steady-state value. Therefore, the supplementary disturbance in Fig.~6(a) is actively utilized to compensate for this deviation, whereas the trajectories of supplementary disturbances in Fig.~6(b) remain near the origin, since the scheme with the degraded system data can compensate for the entire model-plant mismatch using only a small amount of supplementary disturbance. From this, we can see the efficacy of the proposed scheme with the data-window strategy under system characteristics change.

\begin{figure}[t!]
\begin{center}
\includegraphics[width=9cm]{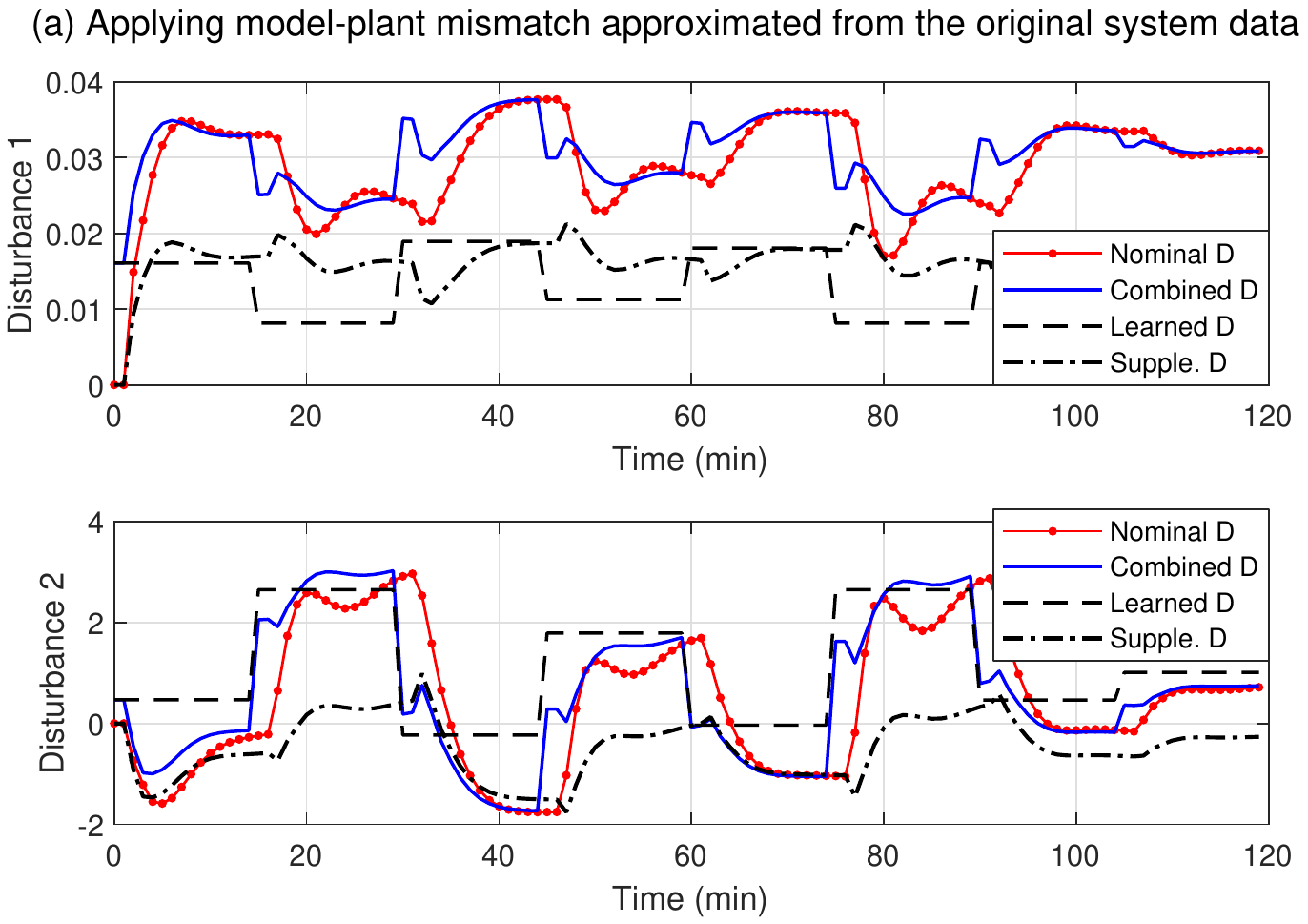}\vspace{0.2cm}\\
\includegraphics[width=9cm]{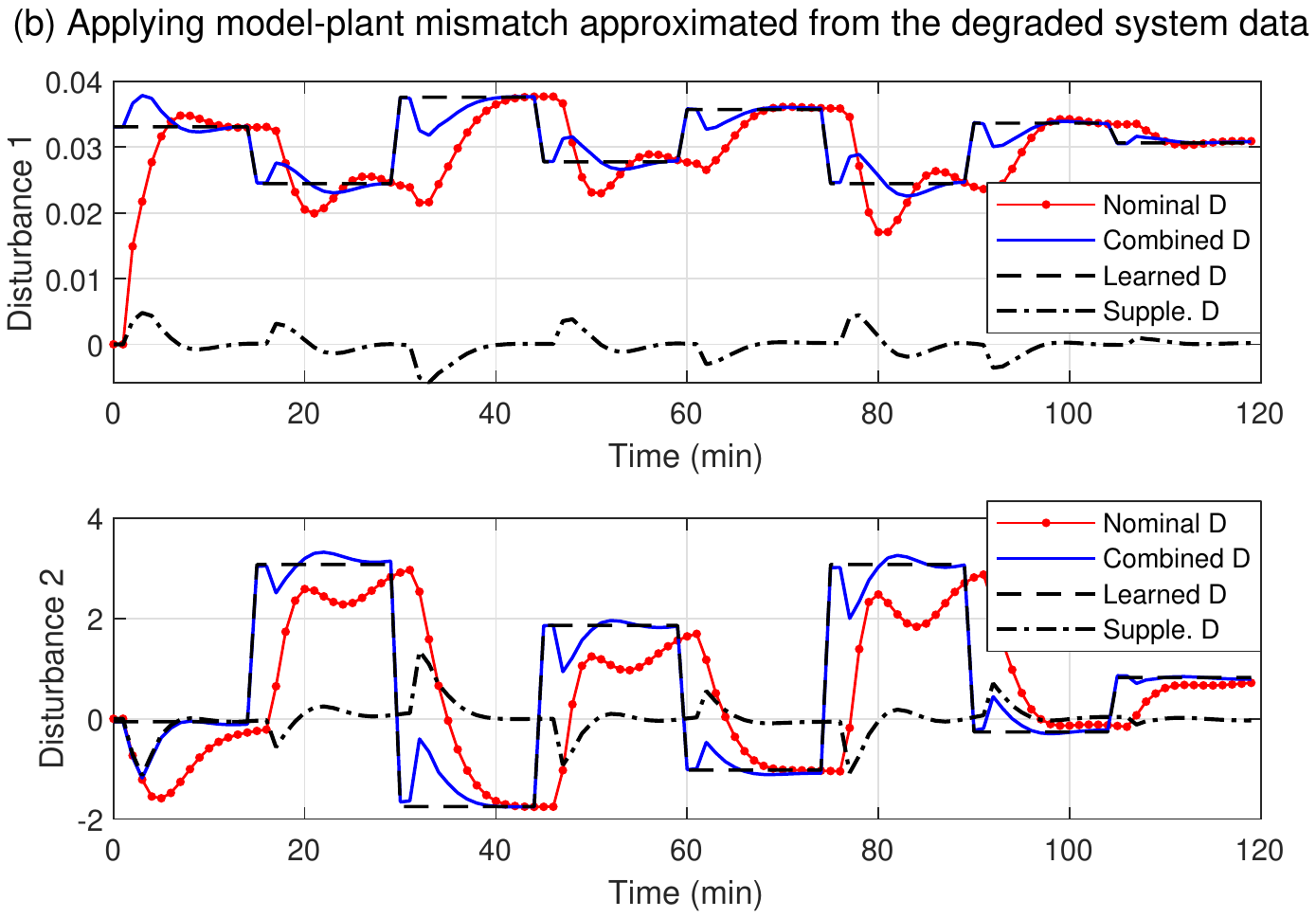}
\caption{Estimated disturbance trajectories of each offset-free MPC scheme with the changed setpoint for $c$ and the fixed setpoint for $T$ by applying model-plant mismatch approximated from (a) the original system data, and (b) the degraded system data.}
\end{center}
\end{figure}

\begin{figure}[t!]
\begin{center}
\includegraphics[width=7cm]{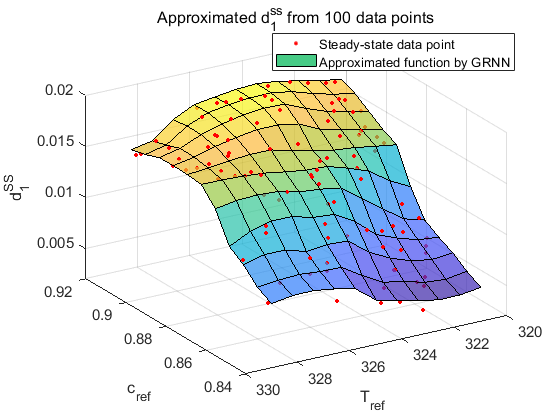}\vspace{-0.1cm}\\
\includegraphics[width=7cm]{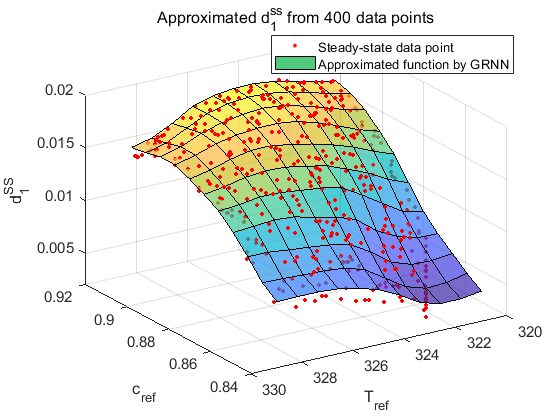}\vspace{-0.1cm}\\
\includegraphics[width=7cm]{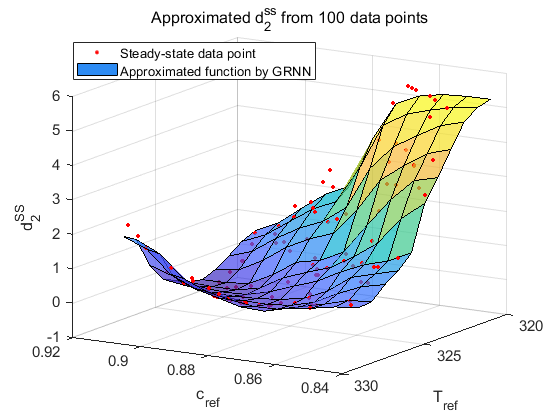}\vspace{-0.1cm}\\
\includegraphics[width=7cm]{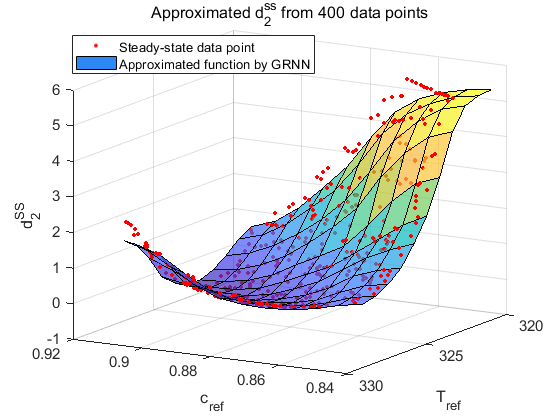}
\caption{Approximated model-plant mismatch by GRNN from 100 and 400 steady-state data points.}
\end{center}
\end{figure}

\subsection{System with setpoint changes for multiple variables}

In this section, we examine the closed-loop behavior of the proposed model-plant mismatch learning offset-free MPC in the system with multiple setpoint changes.

Fig.~7 illustrates the approximation result of the steady-state model-plant mismatch $f_d$ by GRNN. Since the steady-state disturbance pair ($d_1^{ss}$,$d_2^{ss}$) depends on each setpoint pair ($c_{ref}$,$T_{ref}$), the approximated maps for $d_1^{ss}$ and $d_2^{ss}$ are illustrated separately. We can see that the proposed scheme properly approximates the model-plant mismatch between (91) and (92), $\hat{f}_d$, using GRNN in the system with multiple setpoint changes.

\begin{figure}[t!]
\begin{center}
\includegraphics[width=9cm]{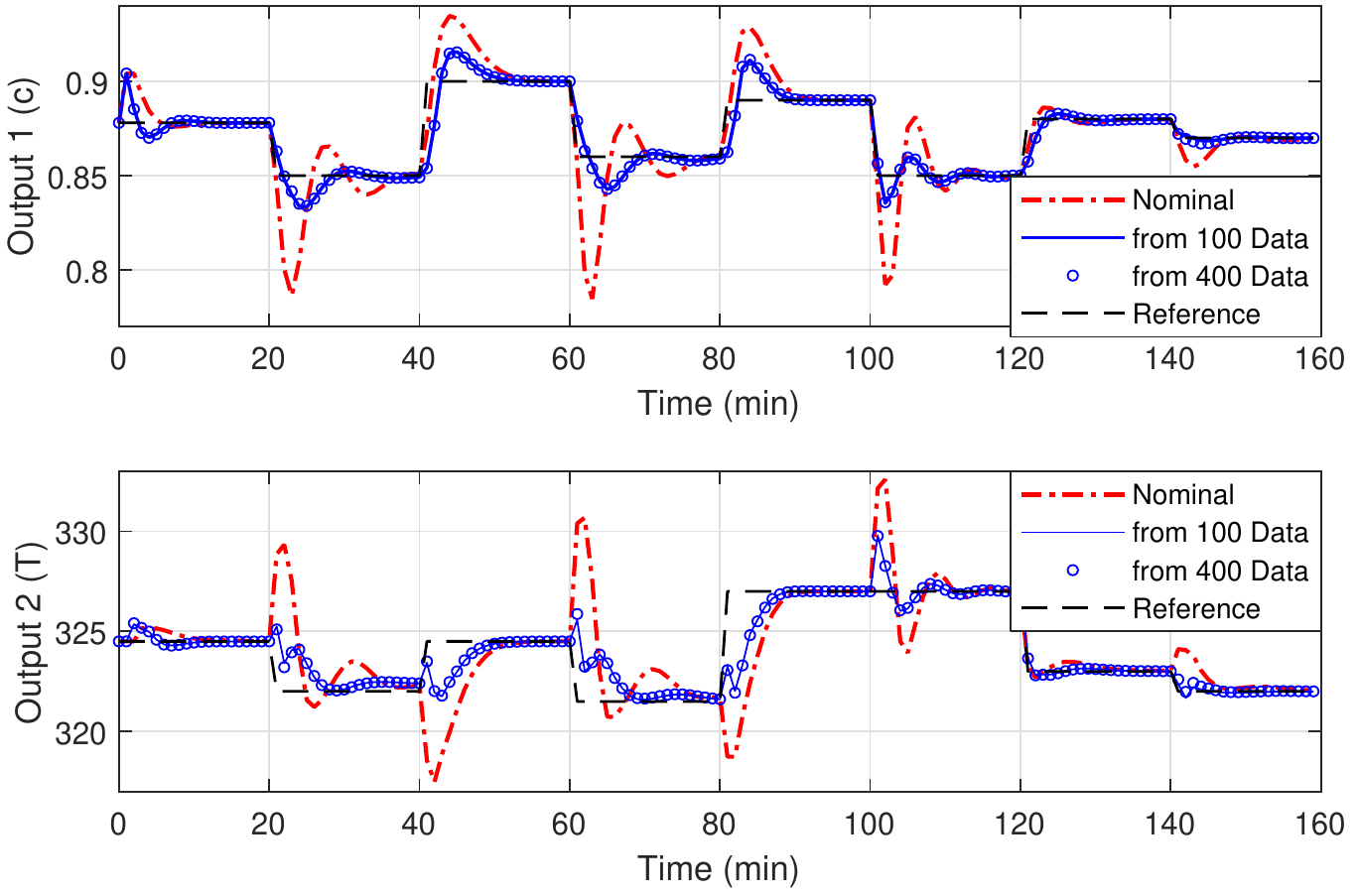}
\caption{Reference tracking results of offset-free MPC schemes with the changed setpoints for $c$ and $T$.}
\end{center}
\end{figure}

Fig.~8 shows the closed-loop simulation results of nominal and proposed offset-free MPC schemes with learned model-plant mismatch from 100 and 400 steady-state data points, respectively, when the setpoints for $c$ and $T$ change along $[0.84,0.91]$ and $[321,329]$ for every 15 min. Both cases utilizing learned model-plant mismatch from 100 and 400 data points show much smaller deviation from the setpoints during transient states than that of the nominal offset-free MPC. Furthermore, although the approximated model-plant mismatch function from 400 data points in Fig.~7 is more exact than that from 100 data points, the closed-loop trajectories of controllers utilizing both approximate functions are almost identical in Fig.~8. From this, we can see that the 100 data points are sufficient to learn a proper approximation of the model-plant mismatch under multiple setpoint changes in this process.

\begin{figure}[t!]
\begin{center}
\includegraphics[width=9cm]{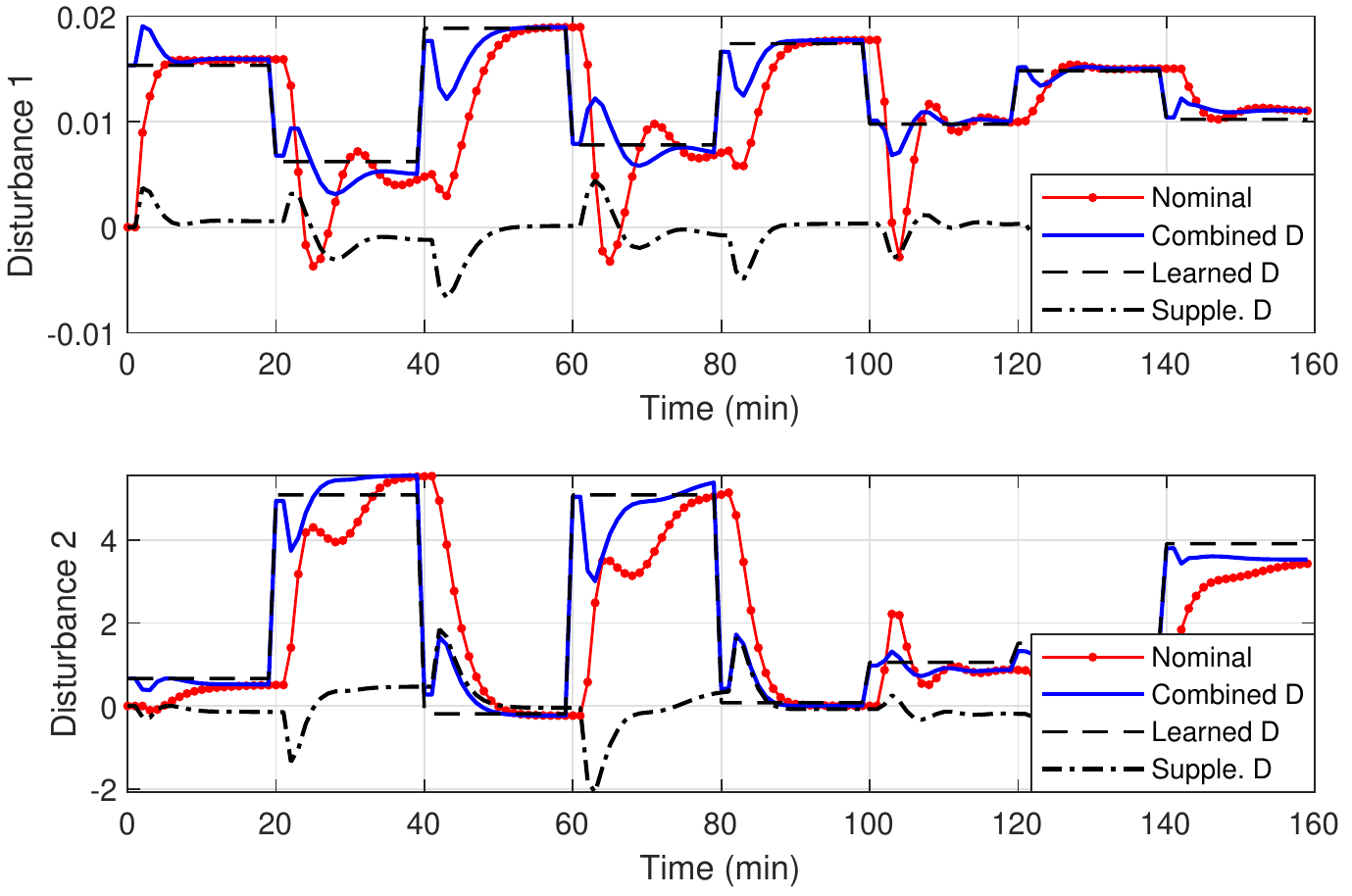}
\caption{Estimated disturbance trajectories of the nominal scheme and the proposed scheme utilizing the learned model-plant mismatch from 100 data points under setpoint changes for $c$ and $T$.}
\end{center}
\end{figure}

Fig.~9 shows the estimated disturbance values of the nominal and the proposed scheme utilizing the learned model-plant mismatch from 100 steady-state disturbance data points in Fig.~7. The learned disturbance values at each setpoint generally match the steady-state disturbance values, but there exist a small quantity of deviations near 40, 80, and 160 min in the second disturbance graph in Fig.~9. This implies that the model-plant mismatch map in Fig.~7 is not perfectly learned near the related setpoints. Even in this case, the proposed scheme achieves the offset-free tracking property and efficiently improves the closed-loop tracking performance, as shown in Fig.~8, by properly utilizing the supplementary disturbance.

\section{Conclusion}\label{sec5}

We developed an improved offset-free model predictive control strategy that performs a predictive control based on a model built on previous knowledge of the process and simultaneously learns the model-plant mismatch from the process data and utilizes it for offset-free tracking with the supplementary signal updated from the disturbance estimator.

With the proposed model-plant mismatch learning offset-free MPC scheme, we effectively improved the closed-loop tracking performance of the controller. In addition, we mathematically analyzed the robust asymptotic stability of the combined system in offset-free MPC consisting of a disturbance estimator, a target problem, and an optimal control problem. Although existing offset-free MPC studies do not examine the closed-loop stability due to the difficulty in considering the combined system, we showed the robust asymptotic stability of the proposed scheme by exploiting the learned model-plant mismatch information.

Lastly, since we combined data-driven and model-based control schemes based on offset-free MPC, the proposed scheme can utilize the model-plant mismatch compensating property of the disturbance model and estimator, which is not available with the nominal MPC. This is the main reason why the proposed method can effectively improve the closed-loop tracking performance without enormous data, unlike existing schemes that improve model-based control performance by updating entire dynamics or learning the entire model-plant mismatch compensating signal directly. Thus, in conclusion, the proposed model-plant mismatch learning offset-free MPC is expected to present an effective direction for the combination of data-driven and model-based control strategies and provide a theoretical foundation for it.

\section*{Acknowledgements}
This work was supported by the National Research Foundation of Korea (NRF) grant 
funded by the Ministry of Science and ICT (MSIT) of the Korean government (No. 2020R1A2C100550311).

\appendix
\section{Proof of Proposition 4}
We can derive \textbf{Proposition 4} from the definition of uniform continuity.\\
\quad\\
\textbf{Proposition 4}. Let $\Gamma$ compact metric space and $g:\Gamma\rightarrow\mathbb{R}^n$ continuous. Then, there exists a class $\mathcal{K}$ function $\sigma(\cdot)$ such that $|g(p)-g(q)|\leq \sigma(|p-q|)$ for all $p,q\in\Gamma$.\\
\quad\\
\textbf{Proof}. Since $g$ is uniformly continuous on $\Gamma$ by \textit{Theorem 4.19} in \cite*{28}, for every $\varepsilon$, $\delta>0$ exists such that $|g(p)-g(q)|<\varepsilon$ for all $p,q\in\Gamma$ for which $|p-q|<\delta$. Then, we can see that there exists a local overbounding class $\mathcal{K}$ function $\bar{\sigma}(\cdot)$ and $\bar{\delta}>0$ such that $|g(p)-g(q)|\leq \bar{\sigma}(|p-q|)$ for all $p,q\in\Gamma$ for which $|p-q|<\bar{\delta}$ by \textit{Proposition 4} in \cite*{29}. Finally, in a similar manner as in the third part of the proof in \textit{Proposition 20} in \cite{23}, we can find the global overbounding class $\mathcal{K}$ function $\sigma(\cdot)$ such that $|g(p)-g(q)|\leq \sigma(|p-q|)$ for all $p,q\in\Gamma$. \qed

\section{Derivation of (53)}

(53) can be derived by a similar flow of the proof for the input/output-to-state stability of a perturbed linear system in \cite{24}. Since the pair (A,C) is assumed to be observable, there exists an estimator gain $L$ such that $A-LC$ is stable. Then, we can describe the estimated augmented state with $e_{\hat{x}_d}$ in (\ref{eq51}):
\begin{align}\label{b1}
\hat{x}_d^{\ell,s}(k+1)=f_m(\hat{x}_d^{\ell,s}(k),u(k))+LC_{x_d}e_{\hat{x}_d}(k)
\end{align}
where $C_{x_d}:=\begin{bmatrix} C & C_d \end{bmatrix}$.

The exact augmented state can be described with $e_{x_d^+}$ in (\ref{eq49}):
\begin{align}\label{b2}
x_d(k+1)=f_m(x_d(k), u(k))+e_{x_d^+}(k).
\end{align}
Subtracting (\ref{b1}) from (\ref{b2}) yields
\begin{align}\label{b3}
e_{\hat{x}_d}(k+1)=(A_{x_d}-LC_{x_d})e_{\hat{x}_d}(k)+e_{x_d^+}(k).
\end{align}
Then, we can describe $e_{\hat{x}_d}(k)$ as in (\ref{b4}) from (\ref{b3}).
\begin{align}\label{b4}
e_{\hat{x}_d}(k)=&(A_{x_d}-LC_{x_d})^k e_{\hat{x}_d}(0) \nonumber\\ &+\sum_{j=0}^{k-1}(A_{x_d}-LC_{x_d})^{k-j-1}e_{x_d^+}(j).
\end{align}

Since $A_{x_d}-LC_{x_d}$ is stable, there exists $\bar{\lambda}$ that satisfies
\begin{subequations}
\begin{align}
&|(A_{x_d}-LC_{x_d})^n|\leq c\bar{\lambda}^n \label{b5a}\\
&\lambda_{max} < \bar{\lambda} <1 \label{b5b}
\end{align}
\end{subequations}
where $\lambda_{max}$ is the maximum eigenvalue of $A_{x_d}-LC_{x_d}$ \citep*{36}.

Then, taking the norm of both sides of (\ref{b4}) and applying (\ref{b5a}) yields
\begin{align}\label{b6}
|e_{\hat{x}_d}(k)|\leq c\bar{\lambda}^k |e_{\hat{x}_d}(0)|+ \frac{c}{1-\bar{\lambda}} ||\mathbf{e}_{x_d^+}||
\end{align}
where $||\mathbf{e}_{x_d^+}||$ denotes the supremum norm of $\mathbf{e}_{x_d^+}$, and $\mathbf{e}_{x_d^+}$ denotes the sequence of $e_{x_d^+}$.

Since $\beta_{\hat{x}_d}(w,k):=c\bar{\lambda}^kw$ is a class $\mathcal{KL}$ function and $\sigma_{\hat{x}_d}(w):=\frac{c}{1-\bar{\lambda}} w$ is a class $\mathcal{K}$ function, we can describe (\ref{b6}) as (\ref{b7}).
\begin{align}\label{b7}
|e_{\hat{x}_d}(k)|\leq \beta_{\hat{x}_d}(|e_{\hat{x}_d}(0)|,k)+\sigma_{\hat{x}_d}(||\mathbf{e}_{x_d^+}||).
\end{align}

\section{Derivation of (55)}

We define the target difference $e_{\bar{x},\bar{u}}$ between the ideal target pair $(\bar{x}^{\ell^*},\bar{u}^{\ell^*})$ and the derived target pair $(\bar{x}^{\ell,s},\bar{u}^{\ell,s})$ from the state and disturbance estimates as (\ref{c1}).
\begin{align}\label{c1}
e_{\bar{x},\bar{u}}:=\begin{bmatrix}{\bar{x}}^{\ell^*}\\ \bar{u}^{\ell^*}\end{bmatrix}-\begin{bmatrix}\bar{x}^{\ell,s}\\ \bar{u}^{\ell,s}\end{bmatrix}.
\end{align}
In the proposed scheme, $\hat{d}^{\ell}$ acts as a base signal for each $\bar{r}$ in the disturbance estimator. Therefore, since the completely learned $\hat{d}^{\ell^*}$ can guarantee considerable prediction accuracy, we assume that supplementary signal $\hat{d}^s$ is bounded (we can see that $\hat{d}^s$ is actually bounded near the origin in numerical examples in Section~\ref{sec4}):
\begin{align}\label{c2}
|\hat{d}^s|\leq c_{\hat{d}^s}.
\end{align}

From (31), (\ref{c3}) holds.
\begin{align}\label{c3}
\begin{bmatrix} -B_d \hat{d}^{\ell,s} \\ \bar{r}-HC_d \hat{d}^{\ell,s} \end{bmatrix}=\begin{bmatrix} -B_d \hat{d}^{\ell^*} \\ \bar{r}-HC_d \hat{d}^{\ell^*} \end{bmatrix}+\begin{bmatrix} -B_d \hat{d}^{s} \\ -HC_d \hat{d}^{s} \end{bmatrix}.
\end{align}
Then, we can derive (\ref{c4}) by substituting (7), (46) and (\ref{c1}) into (\ref{c3}) and rearranging.
\begin{align}\label{c4}
\begin{bmatrix} A-I & B \\ HC & 0 \end{bmatrix} e_{\bar{x},\bar{u}} = \begin{bmatrix} B_d \\ HC_d  \end{bmatrix} \hat{d}^{s}.
\end{align}
By substituting (\ref{c4}) into (\ref{c2}), we can see $e_{\bar{x},\bar{u}}$ is also bounded:
\begin{align}\label{c5}
|e_{\bar{x},\bar{u}}|\leq \left| \begin{bmatrix} A-I & B \\ HC & 0 \end{bmatrix}^{\dagger} \begin{bmatrix} B_d \\ HC_d  \end{bmatrix}\right| c_{\hat{d}^{s}}.
\end{align}
where $\dagger$ represents the pseudoinverse and $|M|$ represents the matrix norm of $M$ corresponding to the Euclidean norm for vectors.

Now, we examine the influence of $e_{\bar{x},\bar{u}}$ on the closed-loop behavior of the optimal control problem. We can reformulate the objective function in P and P$_{\ell^*}$ as the quadratic function for the input sequence $\mathbf{u}:=[u_0^\top,\cdots,u_{N-1}^\top]^\top$ by directly substituting $x_{k+i|k}=A^ix_k+\sum^{i-1}_{j=0} A^j(Bu_{k+i-1-j}+B_dd_k)$:

\begin{equation}\label{c6}
J(\hat{x}_d,\mathbf{u})=\mathbf{u}^\top H_j\mathbf{u} +2\mathbf{u}^\top f_j+c_j
\end{equation}
where
\begin{align}
&H_j:=\Psi ^\top Q_{\mathbf{x}}\Psi +Q_{\mathbf{u}}\nonumber\\
&f_j:=\Psi ^\top Q_{\mathbf{x}} (\Phi \hat{x}+\Psi_{\mathbf{d}} \hat{\mathbf{d}}-\bar{\mathbf{x}})-Q_\mathbf{u}\bar{\mathbf{u}} \nonumber\\
&c_j:=||\Phi \hat{x}+\Psi_\mathbf{d} \hat{\mathbf{d}}-\bar{\mathbf{x}}||^2_{Q_\mathbf{x}}+||\bar{\mathbf{u}}||^2_{Q_\mathbf{u}}+ ||\hat{x}||^2_{Q_x}\nonumber\\
&\hat{\mathbf{d}}:=\mathbf{1}_N \otimes \hat{d},\;\; \bar{\mathbf{x}}:=\mathbf{1}_N \otimes \bar{x},\;\;\bar{\mathbf{u}}:=\mathbf{1}_N \otimes \bar{u}\nonumber\\
&Q_{\mathbf{x}}:=diag\{ Q_x,\cdots,Q_x,Q_x^N \}\nonumber\\
&Q_{\mathbf{u}}:=diag\{ Q_u,\cdots,Q_u\}\nonumber\\
&\Phi:=\begin{bmatrix} A \\ A^2 \\ \vdots \\ A^{\textit{N}} \end{bmatrix},\; \Psi:=\begin{bmatrix} B & 0 & 0 & \cdots & 0 \\ AB & B & 0 & \cdots & 0 \\ A^2 B & AB & B & \cdots & 0 \\ \vdots & \vdots & \vdots & & \vdots \\ A^{\textit{N}-1}B & A^{\textit{N}-2}B & A^{\textit{N}-3}B & \cdots & B \end{bmatrix} \nonumber\\
&\Psi_{\mathbf{d}}:=\begin{bmatrix} B_d & 0 & 0 & \cdots & 0 \\ AB_d & B_d & 0 & \cdots & 0 \\ A^2 B_d & AB_d & B_d & \cdots & 0 \\ \vdots & \vdots & \vdots & & \vdots \\ A^{\textit{N}-1}B_d & A^{\textit{N}-2}B_d & A^{\textit{N}-3}B_d & \cdots & B_d \end{bmatrix} \nonumber.
\end{align}
The Hessian matrices $H_j$ of $J^0$ and $J_{\ell^*}^0$ are identical, but the difference of the gradient vectors $f_j$ is derived as in (\ref{c7}).
\begin{align}\label{c7}
e_{f_j}=-\begin{bmatrix} \Psi ^\top Q_\mathbf{x} & Q_\mathbf{u}\end{bmatrix}( \mathbf{1}_N\otimes e_{\bar{x},\bar{u}}).
\end{align}
Substituting (\ref{c7}) into (\ref{c5}), we can see $e_{f_j}$ is also bounded:
\begin{align}\label{c8}
&|e_{f_j}|\leq N^{1/2}\left| \begin{bmatrix} \Psi ^\top Q_\mathbf{x} & Q_\mathbf{u}\end{bmatrix} \right| \left| \begin{bmatrix} A-I & B \\ HC & 0 \end{bmatrix}^{\dagger} \begin{bmatrix} B_d \\ HC_d  \end{bmatrix}\right| c_{\hat{d}^{s}}.
\end{align}

Let $\kappa_{p}^{\ell^*}$ and $\kappa_{p}^{\ell,s}$ denote the control laws of $\mathrm{P_{\ell^*}}$ and P, respectively. Then, we can define the difference of the control law as
\begin{align}\label{c9}
e_{\kappa_p}:=\kappa_{p}^{\ell,s}-\kappa_{p}^{\ell^*}.
\end{align}
All the parameters and constraints in P and P$_{\ell^*}$ are identical except $f_j$ in the objective function. Therefore, since the control law of the quadratic program is a continuous piecewise affine function for the gradient vector $f_j$ with fixed $x_d$ \citep*{37}, we can see that there exists a class $\mathcal{K}$ function $\sigma_{\kappa_p}(\cdot)$ satisfying the inequality in (\ref{c10}) by \textbf{Proposition 4}.
\begin{align}\label{c10}
|e_{\kappa_p}|\leq \sigma_{\kappa_p}(|e_{f_j}|).
\end{align}
By substituting (\ref{c8}) into (\ref{c10}), we obtain
\begin{equation}\label{c11}
|e_{\kappa_p}|\leq \sigma_{\kappa_p^d}(c_{\hat{d}^s})
\end{equation}
where
\begin{align}
\sigma_{\kappa_p^d}(z):=&N^{1/2}\left| \begin{bmatrix} \Psi ^\top Q_\mathbf{x} & Q_\mathbf{u}\end{bmatrix} \right| \left| \begin{bmatrix} A-I & B \\ HC & 0 \end{bmatrix}^{\dagger} \begin{bmatrix} B_d \\ HC_d  \end{bmatrix}\right| \sigma_{\kappa_p}(z)\nonumber.
\end{align}

\bibliographystyle{model2-names}
\bibliography{references}

\end{document}